\pdfoutput=1
\documentclass{amsart}
\usepackage{amsmath,amsfonts,bm}
\usepackage{amssymb}
\usepackage{graphicx}
\usepackage{subfigure}

\begin{document}

\title[Acoustic damping in spherical structures]{Damping evaluation for free vibration of spherical structures in elastodynamic-acoustic interaction}

\author{Hady K. Joumaa}

\email{hkj@alum.mit.edu, hady.joumaa@gmail.com}

\subjclass{Primary 74J05, 74H45; Secondary 35Q74}

\keywords{Acoustic damping, Elastodynamics, Acoustic radiation, Spherical structures}

\date{Feb 3rd, 2016}

\begin{abstract}
This paper discusses the free vibration of elastic spherical structures in the presence of an externally unbounded acoustic medium. In this vibration, damping associated with the radiation of energy from the confined solid medium to the surrounding acoustic medium is observed. Evaluating the coupled system response (solid displacement and acoustic pressure) and characterizing the acoustic radiation damping in conjunction with the media properties are the main objectives of this research. In this work, acoustic damping is demonstrated for two problems: the thin spherical shell and the solid sphere. The mathematical approach followed in solving these coupled problems is based on the Laplace transform method. The linear under-damped harmonic oscillator is the reference model for damping estimation. The damping evaluation is performed in frequency as well as in time domains; both investigations lead to identical damping factor expressions.
\end{abstract}
\maketitle
\bibliographystyle{amsplain}
\section{Introduction}

Coupled fluid-solid interaction (FSI) problems are in general complex. They require advanced computational methods to handle them. Various studies have previously investigated the vibration of structures while interacting with a surrounding acoustic medium \cite{Lewis_book, Junger_book, Duffy_paper, Berger_paper, Forrestal_paper, Gottlieb_paper, Mang_paper}. Experimental investigations were also conducted on lightweight aerospace structures to estimate their acoustic radiation damping under a wide band of frequency excitations \cite{Brown_paper}. During this interaction, the structure's energy is continuously radiated into the surrounding fluid medium, resulting in a damped structure's response. In \cite{Wilby_chap}, a general expression for the total loss factor exhibited in the vibration of a solid body when immersed in a fluid medium is given as
   \begin{equation}
  		\eta_\textnormal{t} = \eta_{\textnormal{struc}} + \eta_{\textnormal{aero}} + \eta_{\textnormal{rad}}
    \end{equation}
The total loss factor $\eta_\textnormal{t}$, consists of three components: the structural loss factor $\eta_{\textnormal{struc}}$, which represents damping associated with the intrinsic material properties of the structure (e.g., viscoelasticity), the aerodynamic loss factor $\eta_{\textnormal{aero}}$, which is due to the presence of a non-zero mean flow over the structure, and finally, the radiation loss factor $\eta_{\textnormal{rad}}$, the focus of our work, which results from the radiation of sound as a consequence of the structure's vibration. Suppressing the structural loss by assuming a non-dissipative material model and eliminating the aerodynamic loss by considering a perturbing flow in a quiescent fluid medium, the acoustic radiation becomes the sole source of damping  pertaining to the problem. Obviously, when the structure is vibrating in vacuum, no acoustic radiation loss is encountered and thus, $\eta_{\textnormal{rad}} = 0$. In many situations, acoustic radiation effects can be reasonably neglected; however, in some cases, particularly for thin lightweight structures, acoustic damping can be an order of magnitude higher than its structural counterpart \cite{Brown_paper}. The major objective of this research is to formulate and validate a closed-form expression for the acoustic radiation damping factor ($\eta_{\textnormal{rad}}$) revealing its dependence on the physical parameters of the coupled problem. The development of this mathematical expression constitutes the real novelty of our work since no such evaluation has been conducted in previous acoustic radiation damping investigations.
 
\begin{figure}
      \centering
      \scalebox{0.7}
     {
      \includegraphics{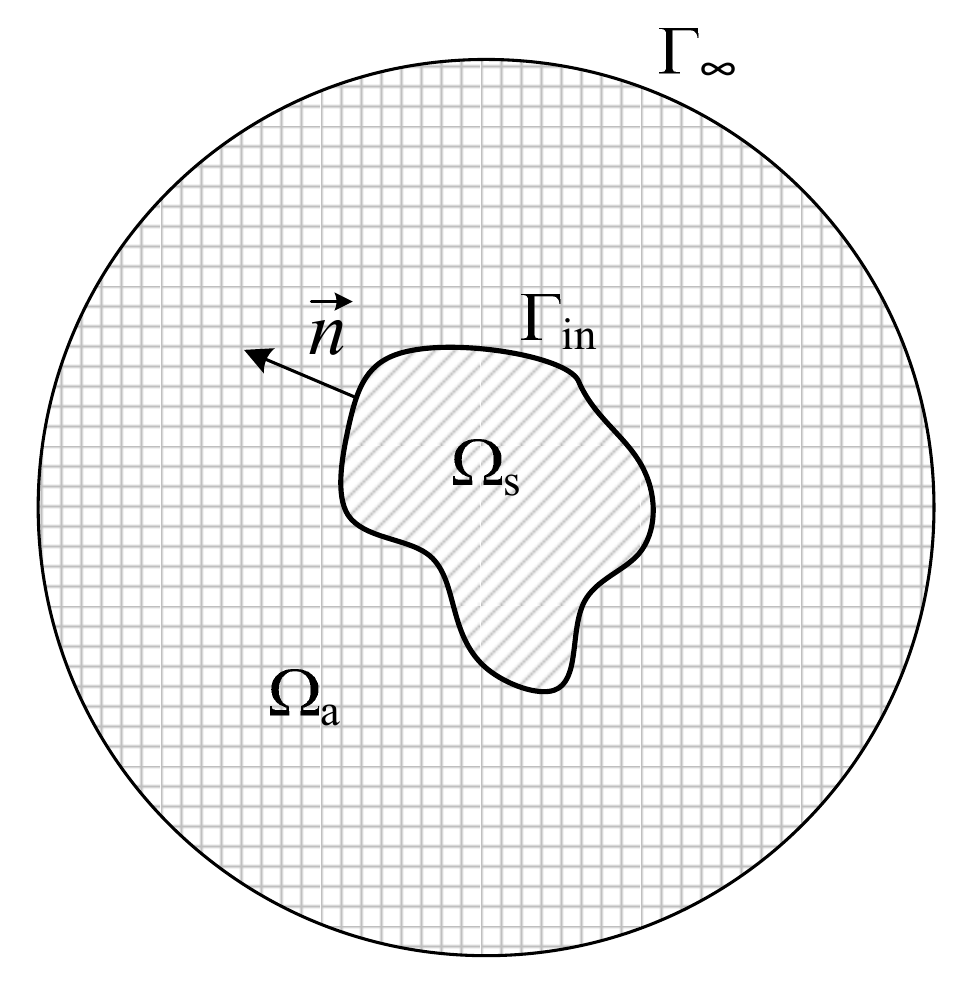}
      }
      \caption{Schematic layout for the general acoustic-structure problem with shown outward normal vector on the common interface}
      \label{schematiclayout}
      \end{figure}

The generalized acoustic-structure interaction problem, whose schematic layout is shown in Figure \ref{schematiclayout}, is comprised of a solid medium $\Omega_{\textnormal{s}}$ that is wholly immersed in an acoustic medium $\Omega_{\textnormal{a}}$. For our work, we adopt the simplest possible models for both $\Omega_{\textnormal{s}}$ and $\Omega_{\textnormal{a}}$ in hope of making analytical solutions feasible. Therefore, the small-strain Hookean model whose elastodynamics is governed by the Navier equation \cite{Graff_book}, and the inviscid compressible fluid in which the acoustic wave propagation is described by the well-known wave equation \cite{Kinsler_book}, are applied in $\Omega_{\textnormal{s}}$ and $\Omega_{\textnormal{a}}$ respectively. The two media have a common interface $\Gamma_{\textnormal{in}}$ on which essential boundary conditions (BC) are maintained to complete the problem's description. As a physical constraint, no interpenetration is allowed through $\Gamma_{\textnormal{in}}$, thus the normal displacement of the structural medium must equate its fluid counterpart; this ensures a continuous displacement field throughout. In addition, the equilibrium on the common surface requires balancing the solid normal traction to the acoustic pressure. Finally, at the infinite boundary $\Gamma_{\infty}$, the non-reflection (wave absorption) condition is applied to emulate the absence of any reflecting surface within the acoustic medium at far field. In summary, the governing equations that constitute the mathematical core of the general coupled problem are presented as follows
 \begin{subequations}
 \label{Gov_FSI_Eqs}
 \begin{equation}
 \dfrac{1}{1-2\nu}u_{j,ji} +u_{i,jj} =  \dfrac{2(1+\nu)\rho_\textnormal{s}}{E} \ddot{u}_i  \quad \textnormal{in $\Omega_\textnormal{s}$}
 \end{equation}
 \begin{equation}
 p_{,kk} = \dfrac {1}{\vartheta_\textnormal{a}^2} \ddot{p}  \quad \textnormal{in $\Omega_\textnormal{a}$}
 \end{equation}
 \begin{equation}
 \ddot{u}_kn_k = -\dfrac {1}{\rho_\textnormal{a}} p_{,j}n_j  \quad \textnormal{on $\Gamma_{\textnormal{in}}$}
 \end{equation}
 \begin{equation}
 \sigma_{ij}n_j = -pn_i \quad \textnormal{on $\Gamma_{\textnormal{in}}$}
 \end{equation}
 \begin{equation}
 \lim_{\left| r \right| \to \infty} \left| r \right| \left( \dfrac {\partial p}{\partial r} + \dfrac{1}{\vartheta_\textnormal{a}} \dfrac{\partial p}{\partial t} \right) = 0 \quad \textnormal{on $\Gamma_{\infty}$}
 \end{equation}
 \end{subequations}
  where $E$, $\nu$, and $\rho_\textnormal{s}$ are respectively the Young's modulus, Poisson's ratio, and mass density of the solid medium, while $\vartheta_\textnormal{a}$ and $\rho_\textnormal{a}$ are respectively the celerity of sound and mass density corresponding to the acoustic medium. The above model was applied in \cite{Bielak_paper} where an FSI problem is solved computationally. The unknowns of Eq. \eqref{Gov_FSI_Eqs} are, to a great importance, the solid displacement $u$, and less importantly, the acoustic pressure $p$.
In our work, we analytically solve the elastodynamic-acoustic interaction problem described by Eq.~\eqref{Gov_FSI_Eqs} for two structures: the thin spherical shell and the solid sphere. In both problems, the exact solutions for both displacement and pressure fields are presented. Consequently, acoustic damping is demonstrated and the damping factor expression is derived from the frequency domain solution. As for validation, the effects of the media properties, problem geometry, and modal wavenumbers are highlighted through the damping extraction algorithms applied on transient solutions.

In previous works, the acoustic-structure interaction problem was considered in the framework where the dynamics is triggered by an external source of excitation (e.g., applied step load on the structure as in \cite{Duffy_paper} or pressure pulse in the acoustic medium as in \cite{Berger_paper}). In addition, the work presented in \cite{Mang_paper} predicts the acoustic radiation damping of flexible panels where the discrepancy in results corresponding to the uniform/nonuniform pressure theories is revealed. More interestingly, the work of \cite{Gottlieb_paper} treats a spherical structure problem; nevertheless, the approach in solving the problem relies on considering a potential flow in the acoustic part, leading to a rather challenging differential functional equation that was never solved. As a result, no rigorous mathematical expression for the acoustic damping could be produced. The distinction of our work lies in considering the situation of self-excitation, where the initially deformed structure is the sole source of energy pertaining to the coupled system. In such a case, the structure's response has no choice but to decay continuously with time due to the lack of any external energy feed that could compensate for the radiated acoustic energy. In consequence, under self-excitation, the coupled system reveals its \emph{intrinsic} damping characteristics allowing for a direct correlation between the natural response and the acoustic damping behaviour to be established. 
 
The paper first briefly overviews the applied damping extraction methods associated with the transient analysis of the linear under-damped harmonic oscillator. Next, we analyse the thin shell problem, which is characterized by its mathematical simplicity with regard to solution formulation and damping estimation. Then, we discuss the more challenging solid sphere problem in which we present the radial mode of oscillation and its corresponding energy calculations, to later solve the coupled problem and verify the damping expression. Finally, we conduct a qualitative comparison between the radiation damping resistances of our problems and those of some previously analysed problems, to draw meaningful conclusions about general acoustic damping estimation.
\section{Damping extraction methods}
The coupled elastodynamic-acoustic problem is linear as depicted in Eq.~\eqref{Gov_FSI_Eqs}, therefore, the damping behaviour of the coupled system is expected to mimic that of the simplest linear dynamic model: the ideal under-damped harmonic oscillator. This model is widely explored in \cite{Hartog_book} and in other references. Its natural vibration is governed by
\begin{equation}
\label{SDOF_gov_eq}
\ddot{u} + 2\zeta\omega_{\textnormal{n}}\dot{u} + \omega_{\textnormal{n}}^2u = 0
\end{equation}
which admits a normalized displacement solution $u(t)$ given by
\begin{equation}
\label{SDOF_u}
u(t) = \exp(-\zeta\omega_{\textnormal{n}}t)\left[ \cos(\omega_{\textnormal{d}}t) + \frac{\zeta}{\sqrt{1-\zeta^2}}\sin(\omega_{\textnormal{d}}t)\right]
\end{equation}
The total energy stored in the system, $e(t) > 0$, is the sum of kinetic and potential energies. It is expressed in normalized form as
\begin{equation}
\label{SDOF_e}
\begin{split}
e(t) & = u^2 + \frac{\dot{u}^2}{\omega_{\textnormal{n}}^2} \\
& = \exp(-2\zeta\omega_{\textnormal{n}}t)\left[ \frac{1-\zeta^2\cos(2\omega_{\textnormal{d}}t)}{1 - \zeta^2} + \frac{\zeta\sin(2\omega_{\textnormal{d}}t)}{\sqrt{1-\zeta^2}}\right]
\end{split}
\end{equation}
 where the the damped frequency $\omega_{\textnormal{d}}$ is defined as
 \begin{equation}
  \omega_{\textnormal{d}} = \omega_{\textnormal{n}}\sqrt{1 - \zeta^2}
  \end{equation}
  
\begin{figure}
\centering
\subfigure[]
	{
	\includegraphics[scale = 0.6]{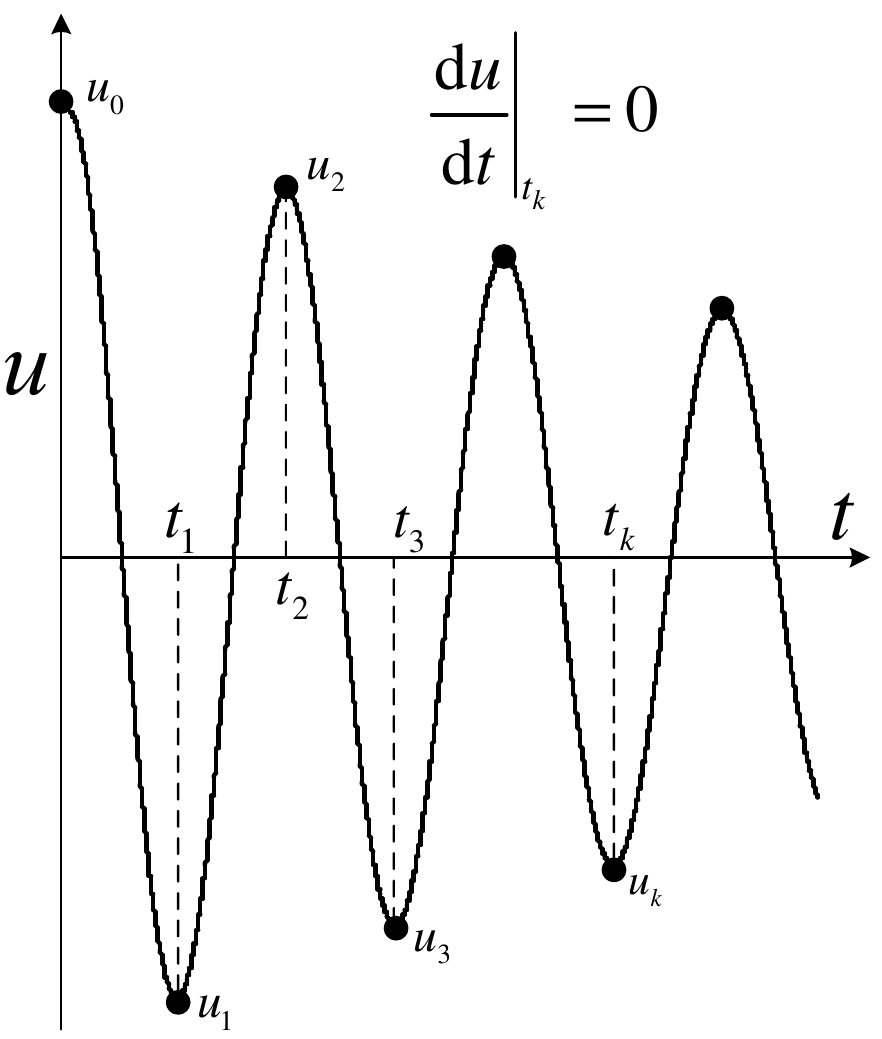}
	\label{disp_response}
}
\subfigure[]
 {
     \includegraphics[scale = 0.6]{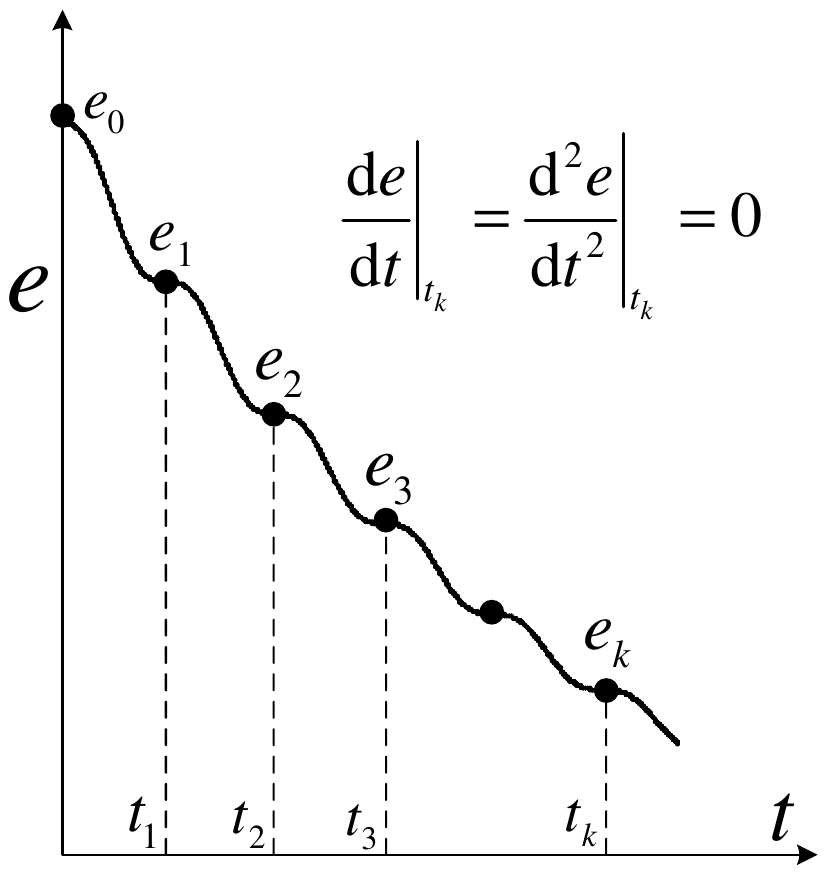}
	 \label{energy_response}
	}
\caption{Transient response for the under-damped harmonic oscillator; labelled are (a): the extrema points of the displacement response and (b): the inflection points of the energy response}
\label{Response Figs_figs}
\end{figure}
In account of the above analysis, mathematical procedures for evaluating the damping property of a dynamic system are established. The log decrement method explained in \cite{Ungar_chap}, is a statistical algorithm that extracts a characteristic damping ratio from the transient response, portraying the intrinsic dissipation of a damped oscillator. Mathematically, for a typical under-damped oscillatory displacement response, consistent with Eq.~\eqref{SDOF_u} and shown in Fig.~\ref{disp_response}, the algorithm predicts an average displacement-based damping ratio, $\bar{\zeta}_{\textnormal{dis}}$, computed as follows
\begin{equation}
\label{log_dec_disp}
\bar{\zeta}_{\textnormal{dis}} = \frac{1}{N}\sum\limits_{n = 0}^{N-1}
\frac{\log \left( \dfrac{ \left| u_{n+1} \right|}{  \left| u_{n} \right| }\right)}
{\sqrt{\log^2 \left( \dfrac{ \left| u_{n+1} \right| }{ \left| u_{n} \right| } \right) + \pi^2}}
\end{equation}
where $u_n$ is the $n$th extremum value of $u(t)$, and $N$ is the total number of all successive extrema used in the evaluation. For the algorithm to generate a meaningful damping ratio, the displacement peak values must steadily decay with time, i.e. $\left| u_{n+1} \right| <  \left| u_{n} \right|$ for all $n$. In some cases, this condition fails (as will be shown in the solid sphere problem), rendering the displacement-based log decrement method erroneous. By necessity to overcome this failure, we extend the applicability of the log decrement method into the energy response, and a corresponding energy-based algorithm is developed. Considering the energy response expressed in Eq.~\eqref{SDOF_e} and plotted in figure \ref{energy_response}, we observe a non-increasing or ``staircase'' behaviour that is typical for this type of dynamic systems. The interesting points to consider for this algorithm are those where the response flats out, i.e. inflection points on which the first and second time derivatives vanish. The average energy-based damping ratio, $\bar{\zeta}_{\textnormal{en}}$, can be evaluated using the following formula
\begin{equation}
\label{log_dec_en}
\bar{\zeta}_{\textnormal{en}} = \frac{1}{N}\sum\limits_{n = 0}^{N-1}
\frac{\log \left( \dfrac{e_{n+1}}{e_n} \right)}
{\sqrt{\log^2 \left( \dfrac{e_{n+1}}{e_n} \right) + 4\pi^2}}
\end{equation}
This algorithm requires the identification of $N$ consecutive inflection points $e_n$, the first one occurs at initial time. The obvious condition for the algorithm's success is the steady decay of energy, i.e. $e_{n+1} < e_{n}$ for all $n$, and this is always satisfied for a damped system subject to natural excitations. Hence, the energy-based log decrement method supersedes its displacement-based counterpart by possessing an unrestricted applicability.

\begin{figure}
\centering
\scalebox{0.45}
{
	\includegraphics{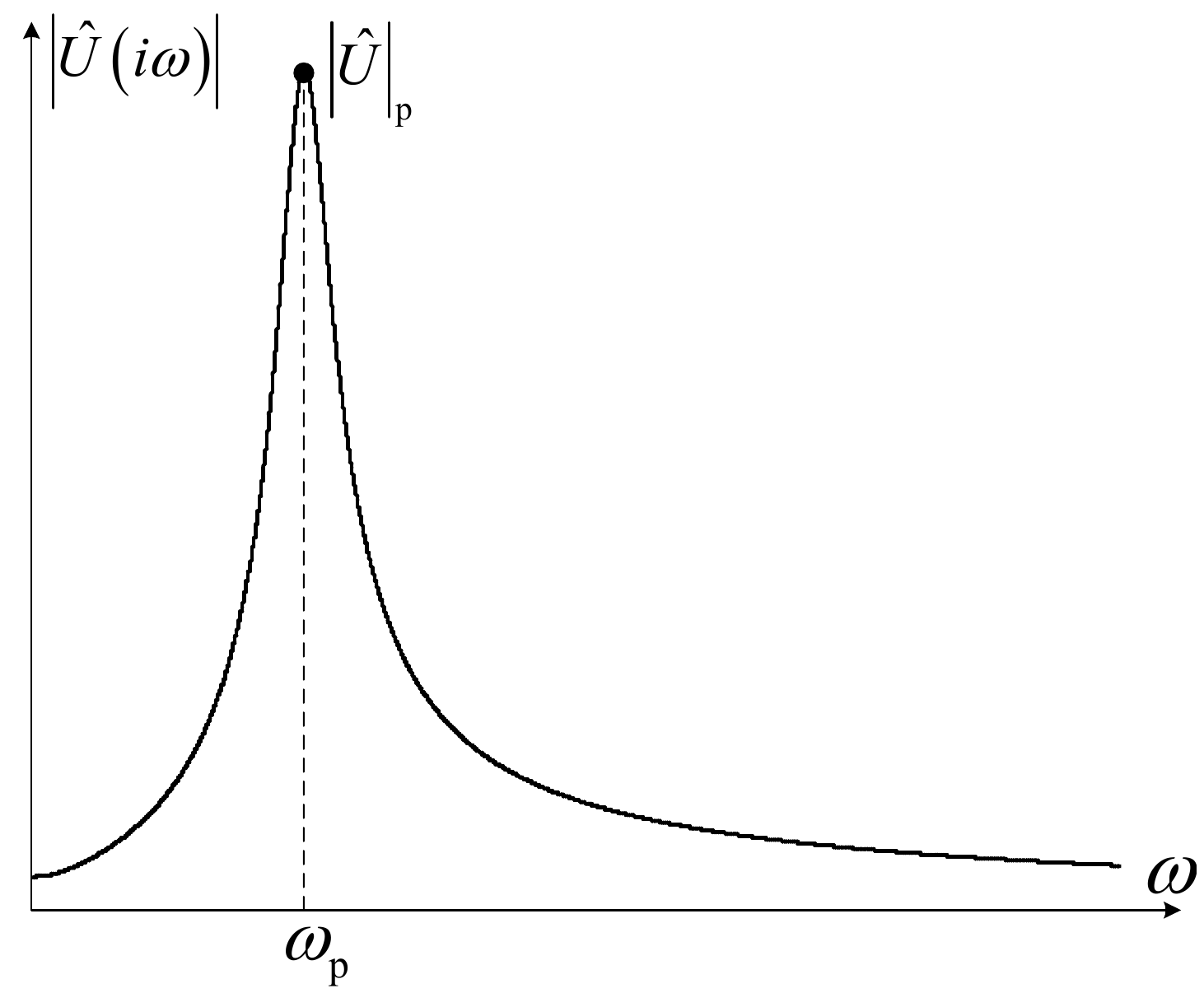}
}
\caption{Frequency response plot for the natural response of the under-damped harmonic oscillator.}
\label{freq_resp_har_osci}
\end{figure}

In addition to damping extraction procedures based on transient analysis, methods based on the frequency (Laplace) response are applied in this work too. Since our analysis is primarily conducted in the frequency domain, the frequency-based damping extraction method generates a meaningful estimate for the damping ratio bypassing all errors involved in the Laplace transform inversion. In addition, this method expresses the damping ratio in closed form by explicitly revealing its physical dependence on the problem's governing parameters. In comparison, the log decrement method simply provides a ``numerical'' value for the damping ratio without unveiling its mathematical structure. In our work, we achieve the damping characterization by first applying the frequency-based method to obtain the closed-form expression of the damping ratio, and then verifying it through log-decrement methods. The mathematical procedure of the frequency method is presented as follows. The Laplace form of $u(t)$ is expressed as
\begin{equation}
\hat{U} \left(s\right) = \dfrac{s + 2\zeta\omega_\textnormal{n}}{s^2 + 2\zeta\omega_\textnormal{n}s + \omega_\textnormal{n}^2}
\end{equation}
The Bode plot of this expression is shown in Fig.~\ref{freq_resp_har_osci}. The response peaks near the natural frequency where
\begin{subequations}
\begin{equation}
\label{omegap_freq}
\omega_\textnormal{p} \cong \omega_\textnormal{n}\left(1 - 4\zeta^4\right)
\end{equation}
\begin{equation}
\label{Up_freq}
\left| \hat{U}\right|_\textnormal{p} \cong \dfrac{1 + 2\zeta^2}{2\zeta\omega_\textnormal{n}}
\end{equation}
\end{subequations}
Inverting Eq.~\eqref{Up_freq} to solve for $\zeta$, we obtain
\begin{equation}
\label{zeta_freqmethod}
\zeta_\textnormal{f} \approx \dfrac{1}{2\sqrt{\left| \hat{U}\right|_\textnormal{p}^2\omega_\textnormal{n}^2 - 1}}
\end{equation}
The above equations are simplified from a more complex form where an approximation for small $\zeta$ is appreciated. Acoustic radiation problems are in general lightly damped; thus the approximation is valid and the analysis is accurate.
  
\section{Pulsating Thin Spherical Shell}
In this section, we present the first and simplest coupled problem: radial pulsation of a thin spherical shell. Consider the elastic thin spherical shell of mean radius $R$ and thickness $b$ where $b\ll R$. The shell's inner core is void and its center resides on the origin of the spherical coordinate system. The spherical symmetry is achieved by considering a uniform and wholly radial displacement field throughout the shell surface, thus $u \equiv u(t)$. This symmetry extends to the surrounding acoustic part, where $p$ spatially depends on the radial distance only, i.e. $p \equiv p(r,t)$. Initially, the acoustic medium is quiescent and the shell is deformed by $u_0$. The governing simplified elastodynamic equation describing the radial oscillation of the shell under this prescription is given as (also applied in \cite{Duffy_paper})
\begin{equation}
\label{shell_elst}
\dfrac{\textnormal{d}^2u}{\textnormal{d}t^2} + \left(\frac{\vartheta_\textnormal{s}}{R}\right)^2 u = -\frac{p_{\textnormal{out}}}{b\rho_\textnormal{s}}
\end{equation}
where $p_{\textnormal{out}}$ is the uniform acoustic pressure applied on the shell's outer surface, i.e. $p\left(R,t\right)$. The structural wave speed is defined as
\begin{equation}
\label{shell_spd}
\vartheta_\textnormal{s} = \sqrt{\frac{2E}{\rho_\textnormal{s}(1-\nu)}}
\end{equation}
On the acoustic side, the acoustic wave equation (D'Alembertian form) expressed in spherical coordinates reduces to
\begin{equation}
\label{shell_wave}
r^2\frac{\partial^2p}{\partial r^2} + 2r\frac{\partial p}{\partial r} - \frac{r^2}{\vartheta_\textnormal{a}^2}\frac{\partial^2p}{\partial t^2} = 0 \; \; \; \; \; \; r \geq R
\end{equation}
On the shell surface, the no-interpenetration condition for small shell radial displacement is expressed as
\begin{equation}
\label{shell_surf}
\frac{\textnormal{d}^2u}{\textnormal{d}t^2} = -\frac{1}{\rho_\textnormal{a}}\frac{\partial p}{\partial r}\arrowvert_{r = R}
\end{equation}
And finally, at far field, the Sommerfeld radiation condition becomes
\begin{equation}
\label{shell_somm}
\lim_{r \to \infty} r\left( \frac {\partial p}{\partial r} + \frac{1}{\vartheta_\textnormal{a}} \frac{\partial p}{\partial t} \right) = 0
\end{equation}
Mathematically, equations \eqref{shell_elst}, \eqref{shell_wave}, \eqref{shell_surf}, and \eqref{shell_somm} describe the coupled problem. The inversion into the Laplace domain is applied to solve for the shell's response. When the wave equation is transformed into the Laplace domain, we obtain the $0{\textnormal{th}}$ order spherical Bessel equation in $r$ admitting the spherical Hankel functions of first and second kind as fundamental solutions. Thus
\begin{equation}
\hat{P}(r,s) = C(s)h_0^{(1)} \left(-\dfrac{isr}{\vartheta_\textnormal{a}}\right) + D(s)h_0^{(2)} \left(-\dfrac{isr}{\vartheta_\textnormal{a}}\right) \; \; \; \; \; \; r \geq R
\end{equation}
where $C(s)$ and $D(s)$ are determined by applying the necessary BC and $i = \sqrt{-1}$. For the Sommerfeld condition in \eqref{shell_somm} to be satisfied, the first kind Hankel function $h_0^{(1)}$ must not appear in the pressure general solution, this is achieved by setting $C(s) \equiv 0$. The reader is advised to review appendix A.4 of \cite{Kinsler_book} or the appendix of \cite{Bender_book} to better understand the asymptotic behaviour of Hankel functions and their derivatives. Substituting for $\hat{P}$ and its derivative in the Laplace versions of Eq.~\eqref{shell_elst} and Eq.~\eqref{shell_surf} and then taking the ratio of these two equations to eliminate $D(s)$, we obtain an expression for $\hat{U}$ given by
\begin{equation}
\dfrac{\hat{U}(s)}{u_0} = \frac{s^2 + s (\frac{\vartheta_\textnormal{a}}{R} + \frac{\vartheta_\textnormal{a}}{b}\frac{\rho_\textnormal{a}}{\rho_\textnormal{s}})}{s^3 + s^2(\frac{\vartheta_\textnormal{a}}{R} + \frac{\vartheta_\textnormal{a}}{b}\frac{\rho_\textnormal{a}}{\rho_\textnormal{s}}) + s\frac{\vartheta_\textnormal{s}^2}{R^2} + \frac{\vartheta_\textnormal{s}^2\vartheta_\textnormal{a}}{R^3}}
\end{equation}
If we normalize the displacement with $u_0$, time with $\frac {R}{\vartheta_\textnormal{s}}$ and consequently the frequency (and Laplace variable, $s$) with $\frac {\vartheta_\textnormal{s}}{R}$, the above expression in normalized form becomes
\begin{equation}
\hat{U}(s) = \frac{s^2 + s (\frac{\vartheta_\textnormal{a}}{\vartheta_\textnormal{s}} + 2\zeta_0)}{s^3 + s^2(\frac{\vartheta_\textnormal{a}}{\vartheta_\textnormal{s}} + 2\zeta_0) + s + \frac{\vartheta_\textnormal{a}}{\vartheta_\textnormal{s}}}
\end{equation}
where $\zeta_0$ is defined as 
\begin{equation}
\label{shell_r_damp_eq}
\zeta_0 = \frac{1}{2} \frac{\rho_\textnormal{a}}{\rho_\textnormal{s}} \frac{\vartheta_\textnormal{a}}{\vartheta_\textnormal{s}} \frac{R}{b}
\end{equation}
Applying the frequency method to extract the damping expression $\zeta_\textnormal{f}$, we obtain
\begin{equation}
\label{shell_freq_damp_eq}
\zeta_\textnormal{f} = \frac{\zeta_0}{\sqrt{1 + 4\zeta_0\frac{\vartheta_\textnormal{a}}{\vartheta_\textnormal{s}}}}
\end{equation}
In real applications, both $\frac{\vartheta_\textnormal{a}}{\vartheta_\textnormal{s}} $ and $\zeta_0 \ll 1$, thus $\zeta_0$ constitutes a first order approximation for $\zeta_\textnormal{f}$. Note that $\zeta_0$ is inversely proportional to $\rho_\textnormal{s}$ and $b$, confirming that acoustic damping intensifies for thin lightweight structures.

\begin{figure}
\centering
\scalebox{0.65}
{\includegraphics{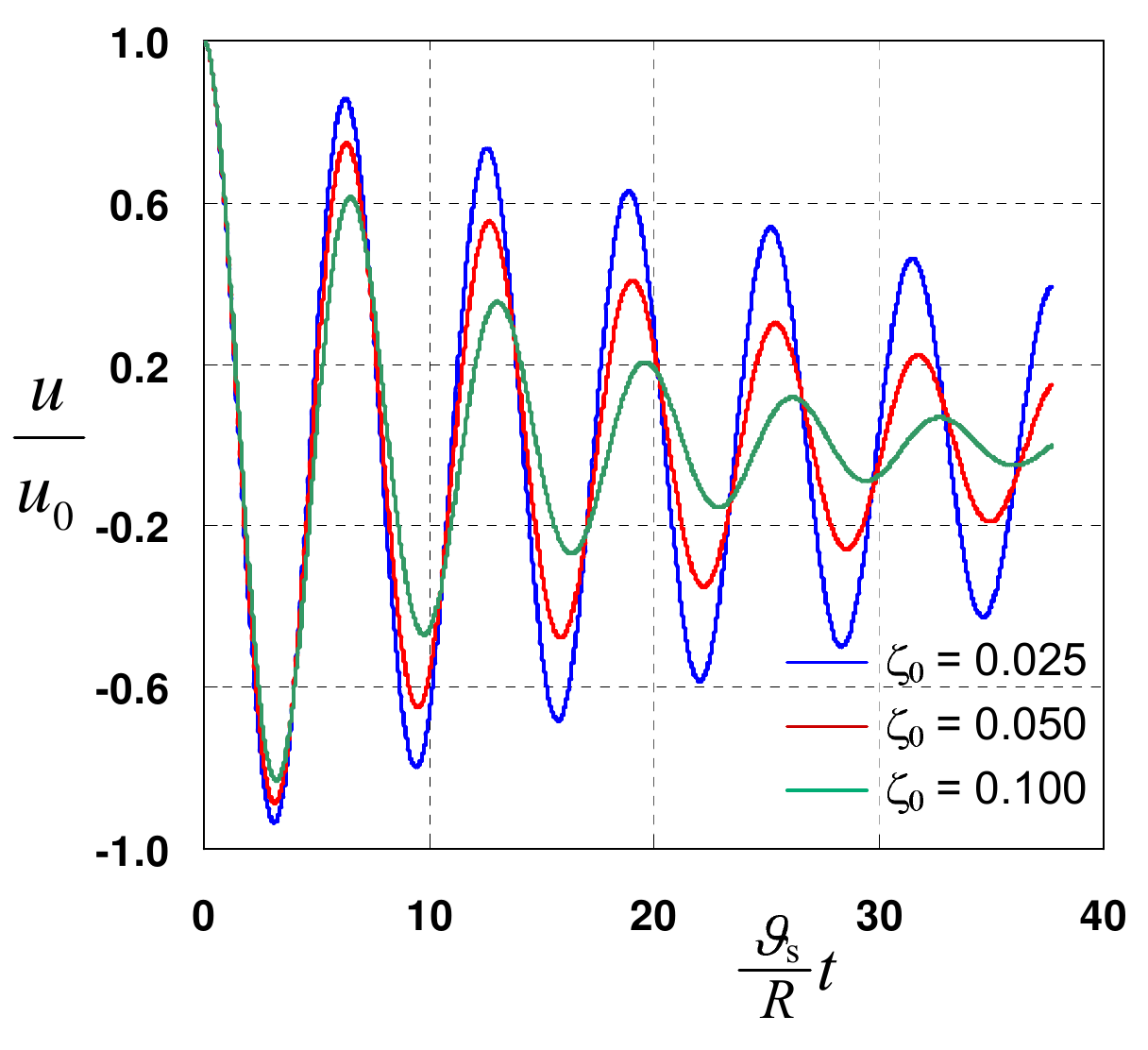}}
\caption{Transient solution for the shell displacement for different $\zeta_0$}
\label{Shell_displacement_plot}
\end{figure}

\begin{figure}
\centering
\scalebox{0.65}
{\includegraphics{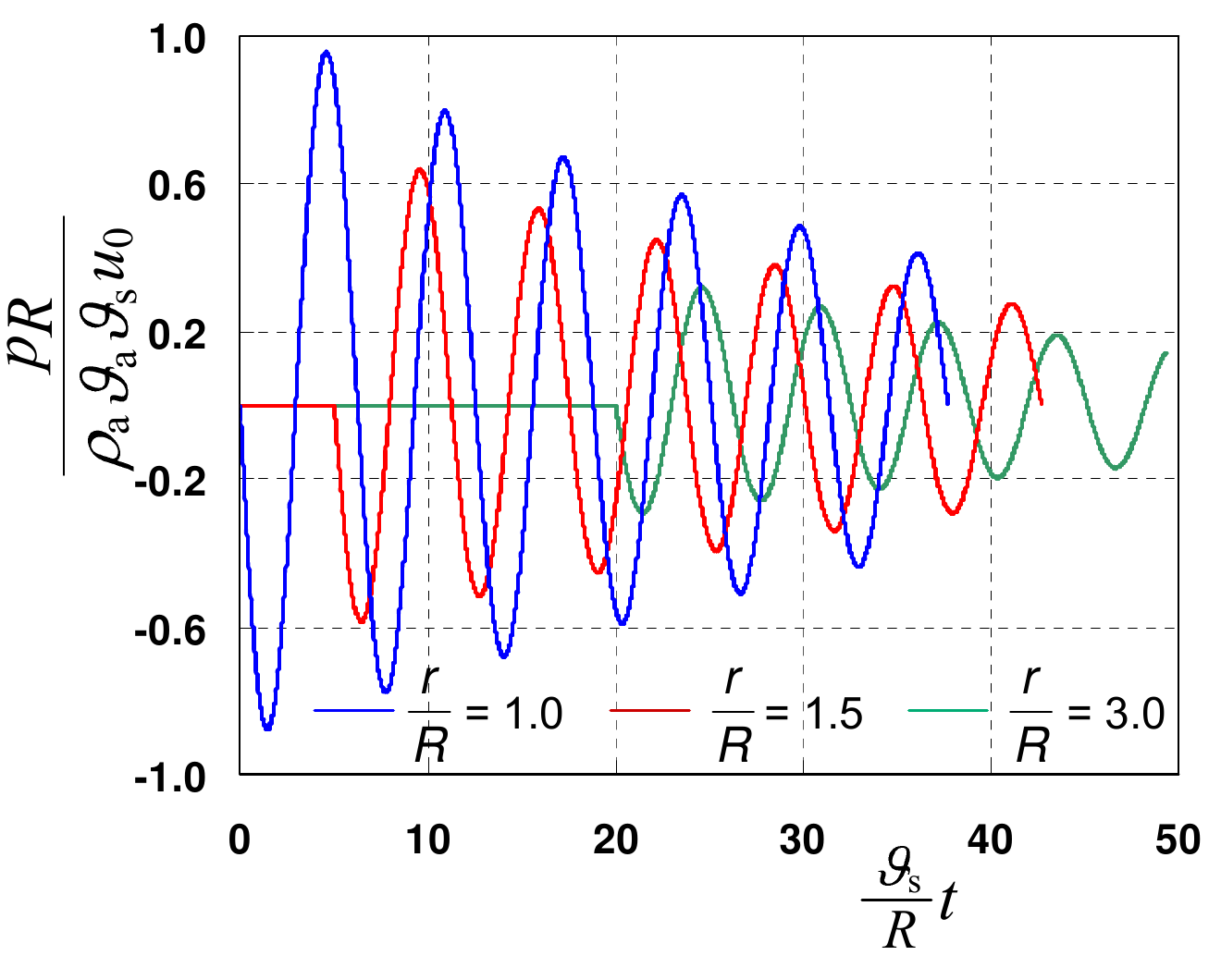}}
\caption{Transient solution for the acoustic pressure at different radial locations for $\zeta_0 = 0.025$ and $\frac{\vartheta_\textnormal{a}}{\vartheta_\textnormal{s}}$ = 0.1}
\label{Shell_pressure_plot}
\end{figure}

Substituting the expression of $\hat{U}$ into the Laplace form of Eq.~\eqref{shell_surf} to solve for $D(s)$, the acoustic pressure (which can be normalized by the characteristic pressure $\rho_\textnormal{a}\vartheta_\textnormal{a}\vartheta_\textnormal{s} \frac{u_0}{R}$) is then obtained as follows
\begin{equation}
\frac{\hat{P}(r,s)}{\rho_\textnormal{a}\vartheta_\textnormal{a}\vartheta_\textnormal{s}\frac{u_0}{R}} = -\frac{R}{r} \frac{s\exp[-\frac{\vartheta_\textnormal{s}}{\vartheta_\textnormal{a}}(\frac{r}{R}-1)s]}{s^3 + s^2(\frac{\vartheta_\textnormal{a}}{\vartheta_\textnormal{s}} + 2\zeta_0) + s + \frac{\vartheta_\textnormal{a}}{\vartheta_\textnormal{s}}}
\end{equation}
Note that both pressure and displacement solutions possess the same characteristic equation (denominator of their Laplace form), thus they exhibit the same dynamic behaviour, in particular, damping behaviour. The Laplace inversion of the displacement and pressure expressions into the time domain is performed via the method of partial fractions (consult chapter 1 of \cite{Cohen_book} and appendix B.2 of \cite{Graff_book} for understanding the inversion technique). Plots of the displacement and pressure solutions are shown in Fig. \ref{Shell_displacement_plot} and Fig. \ref{Shell_pressure_plot} respectively.

Concerning the pressure response, we mark two observations. First, a period of silence i.e. $\left( p = 0\right)$ occurs everywhere in the acoustic domain; its duration increases linearly with the distance from the shell's surface according to
\begin{equation}
T_{\textnormal{silence}} = \frac{r-R}{\vartheta_\textnormal{a}}
\end{equation}
Second, the maximum amplitude of the acoustic pressure, being inversely proportional to the radial distance, decays as we move away from the shell's surface to eventually become zero at infinity. The displacement simulations are conducted at fixed density ratio $\frac{\rho_\textnormal{a}}{\rho_\textnormal{s}} = 0.05$ and fixed geometry $\frac{R}{b} = 10$, but variable wave speed ratio $\frac{\vartheta_\textnormal{a}}{\vartheta_\textnormal{s}} = $ 0.1, 0.2, and 0.4 resulting in three different values for $\zeta_0$ and consequently, three cases 1, 2, and 3 respectively. It is shown that as the ratio $\frac{\vartheta_\textnormal{a}}{\vartheta_\textnormal{s}}$ decreases, the relative error between the transient damping value (obtained by log decrement method) and the closed form value (obtained by frequency method) decreases, and both values approach $\zeta_0$ corroborating the validity of the asymptotic expression of Eq.~\eqref{shell_freq_damp_eq} at low ratio of wave speeds. The results of the three cases are presented in Table ~\ref{shell_table}.
\begin{table}
\caption{Results of the damping factors (in $\%$) for all three cases. Note the strong similarity at low damping.}
\begin{center}
\label{shell_table}
\begin{tabular}{l l l l}
\hline
                                                     & case 1   & case 2     & case 3    \\ \hline
$\zeta _0$                   \%             & 2.500     & 5.000      & 10.000    \\
$\zeta _\textnormal{f}$ \%             & 2.487    & 4.903       & 9.285      \\
$\bar{\zeta} _\textnormal{dis}$ \% & 2.466    & 4.842       & 8.560      \\
Rel. err. \%                                    & 0.851     & 1.260      & 8.470       \\ \hline
\end{tabular}
\end{center}
\end{table}

\section{Vibrating Solid Sphere}
In this section, we demonstrate acoustic damping in the vibrating solid sphere problem which, in concept, coincides with that of the thin spherical shell, but presents some mathematical challenges. The complexity arises because of the presence of modes of vibrations for the solid sphere whose elastodynamics is described by a partial differential equation instead of an ordinary one as is the case for the thin shell. Our first insight about this problem is that the acoustic damping will be mode dependent. Indeed, the closed form expression of the damping factor, along with the results of the transient simulations, corroborate this conjecture. Spherical symmetry will again be enforced to simplify this three-dimensional problem whereby the displacement and the pressure will be solely dependent on the radial distance. Before discussing the coupled problem, we will briefly present the solid sphere radial vibration in vacuum in which we overview the modal analysis and determine the mode shapes and their corresponding natural frequencies. We also introduce the energy calculation which will be applied in the damping evaluation procedures.

\subsection{Elastodynamic overview}
Consider the elastic solid sphere of radius $R$, centered at the origin of the spherical coordinate system. We consider radial modes of vibration, thus the transcendental and the azimuthal displacements are suppressed to zero ($u_{\theta}\equiv 0$, $u_{\phi}\equiv 0$, and $u_{r}\equiv u\left( r,t \right)$). Equations listed in appendix A.9.2 of \cite{Graff_book} help understanding the following analysis. The spherical version of the Navier equation relevant to our problem is
\begin{equation}
\label{sphere_elast_eq}
r^2\frac{\partial^2 u}{\partial r^2} + 2r\frac{\partial u}{\partial r} - 2u = r^2 \frac{\ddot{u}}{\vartheta_\textnormal{s}^2}
\end{equation}
in which the wave speed $\vartheta_\textnormal{s}$ is defined as
\begin{equation}
\label{sphere_spd}
\vartheta_\textnormal{s} = \sqrt{\frac{E(1-\nu)}{\rho_\textnormal{s}(1+\nu)(1-2\nu)}}
\end{equation}
To find the mode shapes $u_\textnormal{m}$, we set $u(r,t)=u_\textnormal{m}(r)e^{i\omega_\textnormal{m}t}$. Substituting this form into Eq.~\eqref{sphere_elast_eq}, the following eigenvalue problem is obtained
\begin{equation}
\label{sphere_modal_eq}
r^2\frac{\textnormal{d}^2 u_\textnormal{m}}{\textnormal{d}r^2} + 2r\frac{\textnormal{d}u_\textnormal{m}}{\textnormal{d}r} + u_\textnormal{m}\left[\left(\frac{r\omega_\textnormal{m}}{\vartheta_\textnormal{s}} \right) ^2 - 2\right] = 0
\end{equation}
This is a first order spherical Bessel equation admitting the spherical Bessel functions of first and second kind as general solutions thus
\begin{equation}
u_\textnormal{m}(r) = C_1j_1\left(\frac{r\omega_\textnormal{m}}{\vartheta_\textnormal{s}}\right) + C_2y_1\left(\frac{r\omega_\textnormal{m}}{\vartheta_\textnormal{s}}\right)
\end{equation}
The BC relevant to this problem are: finite displacement at the origin ($r = 0$) and traction-free outer surface, i.e. $\sigma_{rr}|_{R} = 0$. The first BC forces $C_2$ to vanish because $y_1$ is singular at zero. For simplicity, we set $C_1 = 1$ and we define the normalized modal wave number $\delta_\textnormal{m}$ as
\begin{equation}
\label{sphere_wavenum}
\delta_\textnormal{m} = \frac{\omega_\textnormal{m}R}{\vartheta_\textnormal{s}}
\end{equation}
Hence, the mode shape becomes
\begin{equation}
u_\textnormal{m}(r) = j_1\left(\delta_\textnormal{m}\frac{r}{R} \right)
\end{equation}
\begin{figure}
\centering
\scalebox{0.5}
{
	\includegraphics{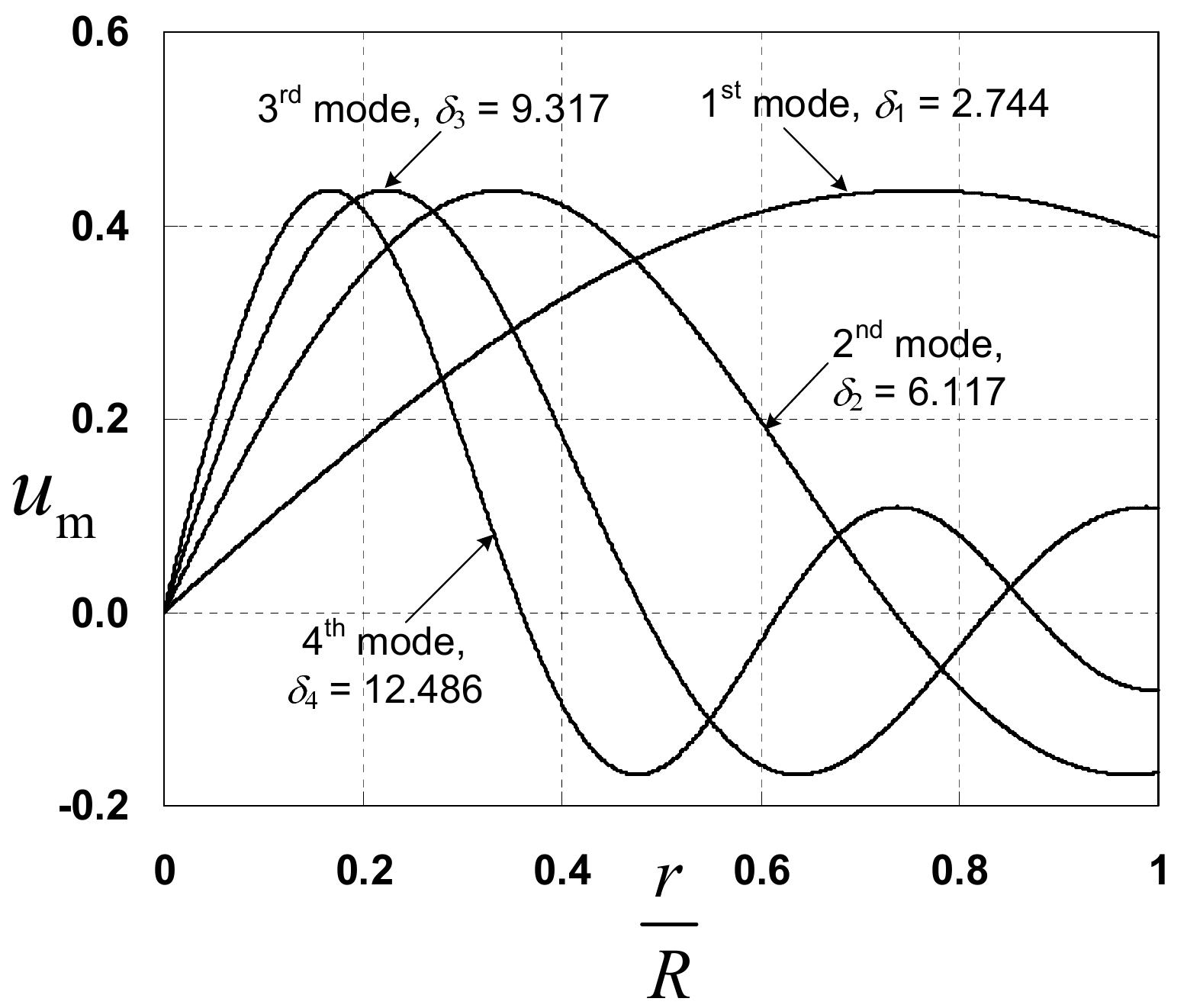}
}
\caption{Modal shapes for the solid sphere radial vibration for $\beta = -1$ with modal wave numbers indicated}
\label{sphere_model_shapes}
\end{figure}
Define a modified Poisson's ratio $\beta$ as follows
\begin{equation}
\beta = \frac{4\nu - 2}{1 - \nu} \quad -2<\beta<0
\end{equation}
The application of the second BC results in the following transcendental equation for $\delta_\textnormal{m}$
\begin{equation}
\tan\delta_\textnormal{m} = \frac{\beta \delta_\textnormal{m}}{\delta_\textnormal{m}^2 + \beta}
\end{equation}
This nonlinear equation is solved numerically via a root finding algorithm (e.g., bisection) to determine the eigenvalues of $\delta_\textnormal{m}$. A meaningful asymptotic approximation for $\delta_\textnormal{m}$ at large integer $m$ is given by $\delta_\textnormal{m} \approx m\pi + \dfrac{\beta}{m\pi}$. The plot in Fig.~\ref{sphere_model_shapes} shows the first four modal functions corresponding to $\beta = -1$ or $\nu = \dfrac{1}{3}$.
\subsection{Energy calculations}
We now introduce the energy calculations corresponding to the radial deformation of this spherical body. We encourage the reader to refer to appendix A.8 and A.9.2 of \cite{Graff_book}
 to better understand the mathematical analysis. For the general case, the stored strain energy per unit volume of an elastic body is defined as
\begin{equation}
e_\textnormal{p} = \int\sigma_{ij}\textnormal{d}\epsilon_{ij}
\end{equation}
In spherical problems, and for the case of radial displacement field depending solely on $r$, the expression for $e_\textnormal{p}$ reduces to
\begin{equation}
e_\textnormal{p} \left(r,t\right) =\frac{\rho_\textnormal{s} \vartheta_\textnormal{s}^2}{2} \left[\left(1 + \frac{\beta}{2}  \right) \epsilon_{ii} \epsilon_{jj} - \frac{\beta}{2} \epsilon_{ij} \epsilon_{ij} \right]
\end{equation}
Clearly, this energy form is space dependent. In order to observe global effects within the entire sphere (example: energy decay), we introduce the volume average strain energy defined as follows
\begin{equation}
\label{average_strain_energy_Eq}
\bar{e}_\textnormal{p}\left(t\right) = \frac{1}{V} \int\limits_{V} e_\textnormal{p}(r,t)\textnormal{d}V = \frac{3}{R^3} \int\limits_{0}^{R}e_\textnormal{p}(r,t)r^2\textnormal{d}r
\end{equation}
In a similar approach, the kinetic energy per unit volume is defined as
\begin{equation}
e_\textnormal{k}\left(r,t\right) = \frac{\rho_\textnormal{s}}{2}\left( \frac{\partial u}{\partial t}\right)^2
\end{equation}
and for the same reason (spatial dependency), we introduce the volume average kinetic energy defined as
\begin{equation}
\label{average_kinetic_energy_Eq}
\bar{e}_\textnormal{k}\left(t\right) = \frac{1}{V} \int\limits_{V} e_\textnormal{k}(r,t)\textnormal{d}V = \frac{3}{R^3} \int\limits_{0}^{R}e_\textnormal{k}(r,t)r^2\textnormal{d}r
\end{equation}
Finally, we define the normalized volume average total energy stored in the solid sphere by summing the strain and kinetic energy and normalizing the resulting sum with its initial value. As a result we have
\begin{equation}
\label{total_energy_norm_sphere}
\bar{e}_\textnormal{t} = \frac{\bar{e}_\textnormal{p} + \bar{e}_\textnormal{k}}{\bar{e}_\textnormal{p}|_{t=0} + \bar{e}_\textnormal{k}|_{t=0}}
\end{equation}
For free vibrations in vacuum, the energy form of \eqref{total_energy_norm_sphere} remains constant (identically one) perpetually. In such a case, an exchange between kinetic and strain energy occurs without any external dissipation or addition of energy. However, in the coupled situation where an interaction between the sphere and its surrounding fluid medium takes place, the total energy decays due to acoustic radiation. The understanding of this decay helps characterizing the acoustic damping, the topic of our investigation.

\subsection{Coupled problem formulation}
In this section, we consider the coupled solid sphere problem, where an acoustic medium fills the entire space surrounding the solid sphere. The goal is to evaluate the acoustic damping corresponding to each mode of vibration. Thus, a single mode is excited in each case study, and the resulting damping factor is recorded. In brief, the trend to obtain a verified expression for the acoustic damping ratio is explained in these steps:
\begin{enumerate}
\item{solve the IBVP (single-mode excitation) and obtain the Laplace form of the displacement and acoustic pressure}
\item{apply the frequency method to extract the closed form expression of the damping ratio}
\item{invert the Laplace form of the displacement into time domain}
\item{evaluate the velocity and strain fields to solve for the total energy}
\item{apply the log decrement methods and verify the results with the expression obtained in step (2)}
\end{enumerate}
For the structural domain, the IBVP is formulated by considering Eq.~\eqref{sphere_elast_eq} with modified BC on the sphere's outer surface. Initially, the sphere is deformed in accordance with the mode of interest; thus the modal function $u_\textnormal{m}(r)$ appears in the Laplace form of Eq.~\eqref{sphere_elast_eq}, which is an inhomogeneous spherical Bessel equation in $r$ expressed as
\begin{equation}
\label{c_sphere_modal_eq}
r^2\frac{\partial^2 \hat{U}}{\partial r^2} + 2r\frac{\partial \hat{U}}{\partial r} + \hat{U}\left[\left(\frac{isr}{\vartheta_\textnormal{s}} \right) ^2 - 2\right] = -\frac{sr^2}{\vartheta_\textnormal{s}^2}u_\textnormal{m}(r)
\end{equation}
The full solution of $\hat{U}$ consists of the homogeneous part (spherical Bessel function) in addition to the particular one imposed by the right-hand side term. The full solution, satisfying the finite displacement BC at the sphere's centre, admits the form
\begin{equation}
\label{cpl_disp_gen_sol}
\hat{U}(r,s) = C(s)j_1\left(\frac{isr}{\vartheta_\textnormal{s}} \right) + u_\textnormal{m}(r)\frac{s}{s^2 + \omega_\textnormal{m}^2}
\end{equation}
Concerning the acoustic medium, the general solution for the pressure is identical to that of the pulsating thin shell problem. This is intuitively explained by observing that the acoustic wave propagation is indifferent to the inner core of the spherical domain whether it is void or filled. As a matter of fact, Eq.~\eqref{shell_wave} and Eq.~\eqref{shell_somm} are independent from any solid influence. Therefore, the acoustic pressure's general solution is given by
\begin{equation}
\label{cpl_pres_gen_sol}
\hat{P}(r,s) = D(s)h_0^{\left( 2 \right)}\left(\frac{-isr}{\vartheta_\textnormal{a}} \right)
\end{equation}
Equations \eqref{cpl_disp_gen_sol} and \eqref{cpl_pres_gen_sol} contain two unknown constants $C(s)$ and $D(s)$; they are determined by applying the coupled BC on the common interface, the sphere's outer surface. The no-interpenetration BC is no different from that of the thin shell problem expressed in Eq.~\eqref{shell_surf}. Rewriting it in Laplace domain, we obtain
 \begin{equation}
  \label{disp_bc_freq}
  \rho_\textnormal{a}\left[s^2\hat{U}\left(R,s\right) - su_\textnormal{m}\left(R\right) \right] = -\frac{\partial \hat{P}}{\partial r}|_R
  \end{equation}
The traction BC balances the radial stress with the acoustic pressure. This BC is expressed in (a) time and (b) Laplace domain as follows
\begin{subequations}
 \begin{equation}
 \label{force_bc_time}
 \sigma_{rr}|_R = -p|_R
 \end{equation}
  \begin{equation}
   \label{force_bc_freq}
 \rho_\textnormal{s}\vartheta_\textnormal{s}^2 \left[ \frac{\partial \hat{U}}{\partial r}|_R + \frac{2\nu}{(1-\nu)R} \hat{U}\left(R,s\right) \right] = -\hat{P}\left(R,s\right)
   \end{equation}
 \end{subequations}
In account of Eq.~\eqref{disp_bc_freq} and Eq.~\eqref{force_bc_freq}, we can solve for $C(s)$ and $D(s)$ after an extensive algebraic work. Hence the Laplace form of the sphere's displacement and the acoustic pressure are obtained.

We introduce the dimensionless parameter $q \equiv \frac{\rho_\textnormal{a}}{\rho_\textnormal{s}} \frac{\vartheta_\textnormal{a}}{\vartheta_\textnormal{s}}$, and we note that $q \ll 1$ for most real applications. This parameter appears in the expression of the damping coefficient as will be shown. We simplify the expressions of the coupled solution by first normalizing the Laplace variable $s$ with $\omega_\textnormal{m}$ and the displacement with $u_\textnormal{m}(R)$, and second, by substituting the spherical Bessel functions of complex argument with their equivalent hyperbolic functions to obtain the solution of the displacement field as follows
\begin{subequations}
\label{U_set_eq}
\begin{equation}
\label{U_tot_eq}
\hat{U}(r,s) = \frac{s}{s^2 + 1} \left[ \frac{\hat{U}_{\textnormal{1,num}}(s)}{\hat{U}_{\textnormal{1,den}}(s)} \cdot \hat{U}_2(r,s) + \frac{u_\textnormal{m}(r)}{u_\textnormal{m}(R)}\right]
\end{equation}
\begin{equation}
\label{U_1_eq}
\hat{U}_{\textnormal{1,num}}(s) = q\delta_\textnormal{m}^2 \left[ \delta_\textnormal{m}s - \tanh \left(\delta_\textnormal{m}s \right) \right]
\end{equation}
\begin{equation}
\begin{split}
& \hat{U}_{\textnormal{1,den}}(s) = q \left(\delta_\textnormal{m} s\right)^3 + \beta\left(\delta_\textnormal{m} s\right)^2 + \beta\frac{\vartheta_\textnormal{a}}{\vartheta_\textnormal{s}}\delta_\textnormal{m} s \\ 
& + \tanh \left(\delta_\textnormal{m}s \right) \left[  \left(\delta_\textnormal{m} s\right)^3 + \left( \frac{\vartheta_\textnormal{a}}{\vartheta_\textnormal{s}} - q \right) \left(\delta_\textnormal{m} s\right)^2
- \beta\delta_\textnormal{m} s - \beta\frac{\vartheta_\textnormal{a}}{\vartheta_\textnormal{s}} \right]
\end{split}
\end{equation}
\begin{equation}
\label{U_2_eq}
\hat{U}_2(r,s) = \frac{R}{r} \frac{\delta_\textnormal{m}s\cosh\left(\frac{r}{R}\delta_\textnormal{m}s \right) -  \frac{R}{r}\sinh\left(\frac{r}{R}\delta_\textnormal{m}s \right) }{\delta_\textnormal{m}s\cosh\left(\delta_\textnormal{m}s \right) - \sinh\left(\delta_\textnormal{m}s \right)}
\end{equation}
\end{subequations}
and that of the acoustic pressure field as
\begin{subequations}
\label{P_set_eq}
\begin{equation}
\label{P_R_eq}
\frac{\hat{P}\left(R,s \right)}{\rho_\textnormal{a} \vartheta_\textnormal{a} \vartheta_\textnormal{s} \frac{u_\textnormal{m}(R)}{R} } =\frac{\delta_\textnormal{m}^2 s}{\frac{\vartheta_\textnormal{a}}{\vartheta_\textnormal{s}} +\delta_\textnormal{m} s} \left[ \frac{s}{s^2 + 1} \left(\hat{U}_1 \left(s \right) + 1 \right) - 1 \right]
\end{equation}
\begin{equation}
\label{P_gen_eq}
\hat{P}\left(r,s \right) = \frac{R}{r} \hat{P}\left(R,s \right) \exp \left[ -\delta_\textnormal{m}\frac{\vartheta_\textnormal{s}}{\vartheta_\textnormal{a}} \left(\frac{r}{R} - 1 \right)s \right]
\end{equation}
\end{subequations}
Clearly, these Laplace expressions are not at all familiar; their inversion technique will be discussed in the upcoming subsection.

\subsection{Damping evaluation}
The frequency method is applied on the Laplace form of the displacement presented in Eq.~\eqref{U_set_eq}. On the outer surface $(r = R)$, $\hat{U}_2 \equiv 1$ and the expression of $\hat{U}$ reduces to
\begin{equation}
\label{U_outer_eq}
\hat{U}(R,s) = \frac{s}{s^2 + 1} \left[ \hat{U}_1(s) + 1\right]
\end{equation}
\begin{figure}
\centering
\scalebox{0.7}
{
	\includegraphics{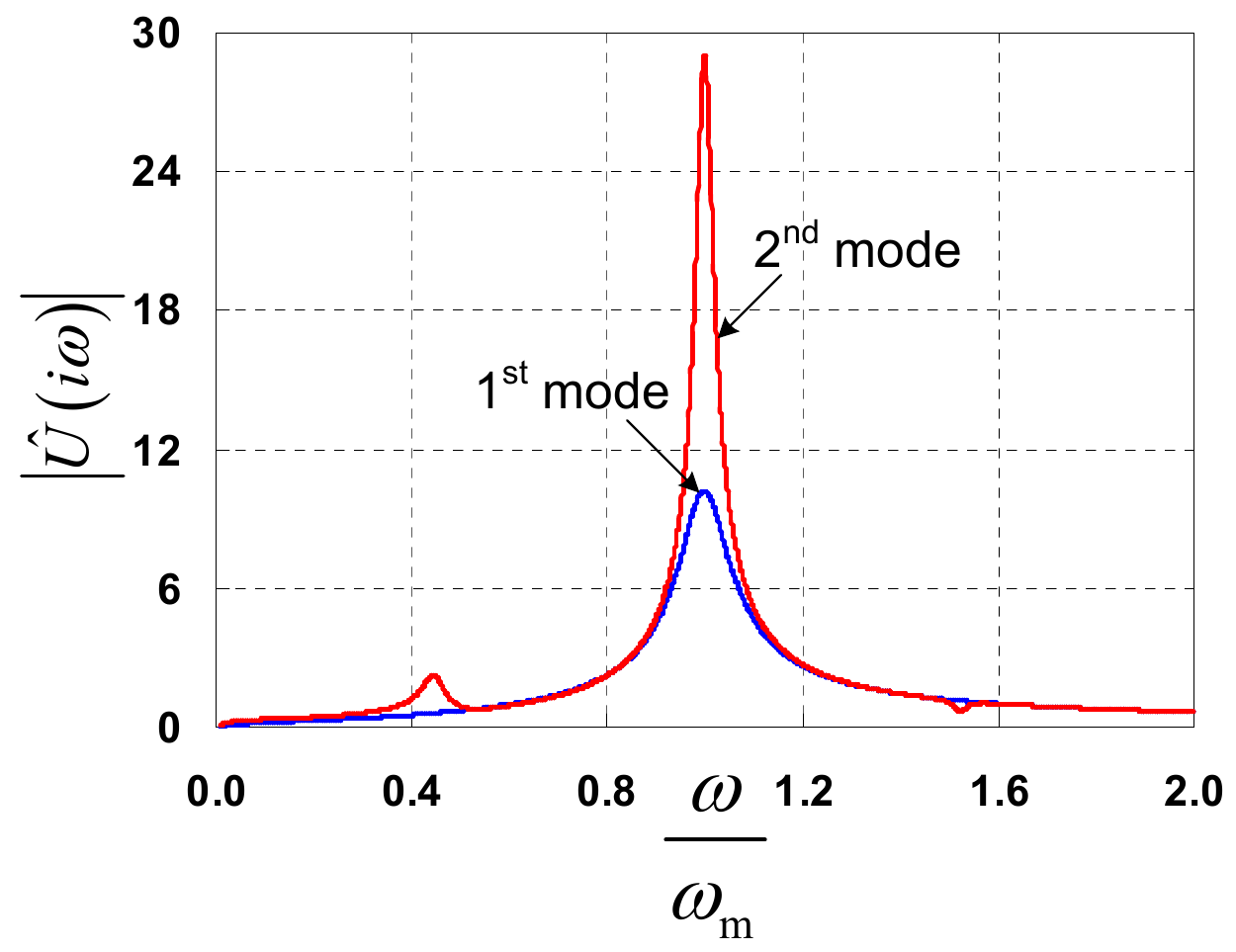}
}
\caption{Bode plot of the sphere's outer surface displacement corresponding to the first two modal excitations}
\label{Bode_sphere_plot}
\end{figure}
The amplitude of $\hat{U}(R,s)$ is plotted in Fig.~\ref{Bode_sphere_plot}. Note that this frequency response is behaviourally identical to that of the harmonic oscillator shown in Fig.~\ref{freq_resp_har_osci}. The major peak of the response occurs at frequency slightly below the natural frequency ($\omega_{\textnormal{d}} \approx \omega_{\textnormal{m}}$). The value of the peak increases as we go higher in modes, hinting to damping attenuation for higher modes. In addition, second and higher modes exhibit secondary peaks occurring far from the frequency of excitation; these peaks are irrelevant to damping evaluation but affect the transient response as will be realized. Solving analytically for the peak frequency and the corresponding response amplitude is not feasible due to the mathematical complexity involved in the $\hat{U}(R,s)$ expression. We will approximate the peak of the amplitude response by the value occurring at the natural frequency, in other words
\begin{equation}
\label{U_approx_freq}
\begin{split}
\left|\hat{U}_\textnormal{p}\right|  \cong & \left|\lim_{s \to i} \hat{U}(R,s)\right| \\ = & 
\left| \frac{\delta_\textnormal{m}^2 + \beta^2 + 3\beta}{2q\delta_\textnormal{m}} -
 i \left( 1 + \frac{\vartheta_\textnormal{a}}{\vartheta_\textnormal{s}}\frac{\delta_\textnormal{m}^2 + \beta^2 + 3\beta}{2q\delta_\textnormal{m}^2} \right) \right|
\end{split}
\end{equation}
The damping value can be then obtained using Eq. \eqref{zeta_freqmethod}, thus
 \begin{equation}
 \label{zeta_m_freq}
 \zeta_{\textnormal{m,f}} \cong \frac{q\delta_\textnormal{m}}{\delta_\textnormal{m}^2 + \beta^2 + 3\beta} \frac{1}{\sqrt{1 + \left( \frac{\vartheta_\textnormal{a}}{\delta_\textnormal{m} \vartheta_\textnormal{s}} \right)^2 + 4\frac{q}{\delta_\textnormal{m}^2 + \beta^2 + 3\beta} \frac{\vartheta_\textnormal{a}}{\vartheta_\textnormal{s}} }}
 \end{equation}
 Note that the square root term in Eq. \eqref{zeta_m_freq} can be safely disregarded for small $\frac{\vartheta_\textnormal{a}}{\vartheta_\textnormal{s}}$ and high modes (large $\delta_\textnormal{m}$). In spite of the approximation applied to obtain the final form of $\zeta_{\textnormal{m,f}}$, the latter still meaningfully evaluates the acoustic damping as is confirmed by the transient analysis results shown in Table ~\ref{Damping_Table}.

\begin{figure}
\centering
\subfigure[first mode]
	{
	\includegraphics[scale = 0.43]{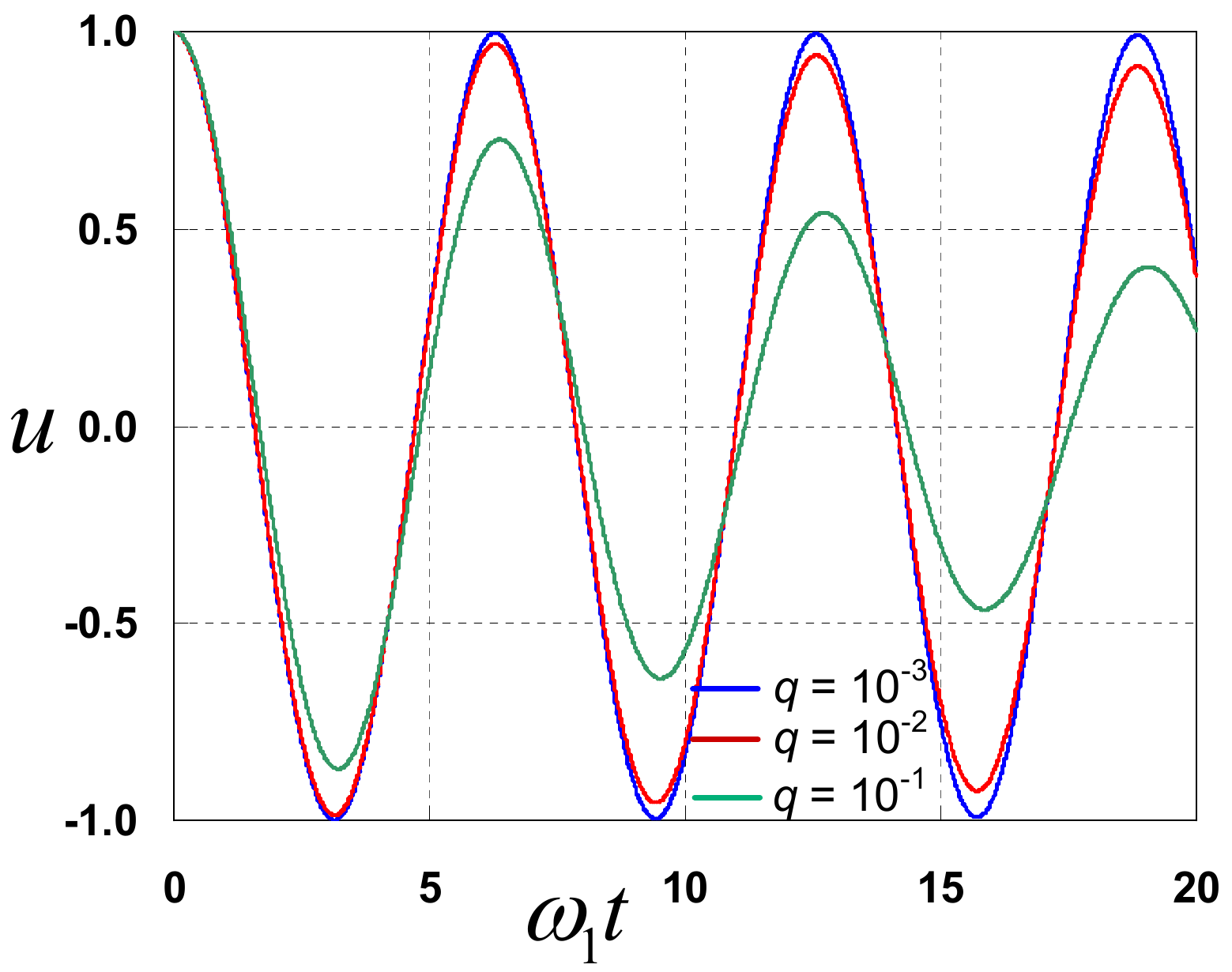}
	\label{disp_sphere_mode_1}
}
\subfigure[second mode]
 {
     \includegraphics[scale = 0.43]{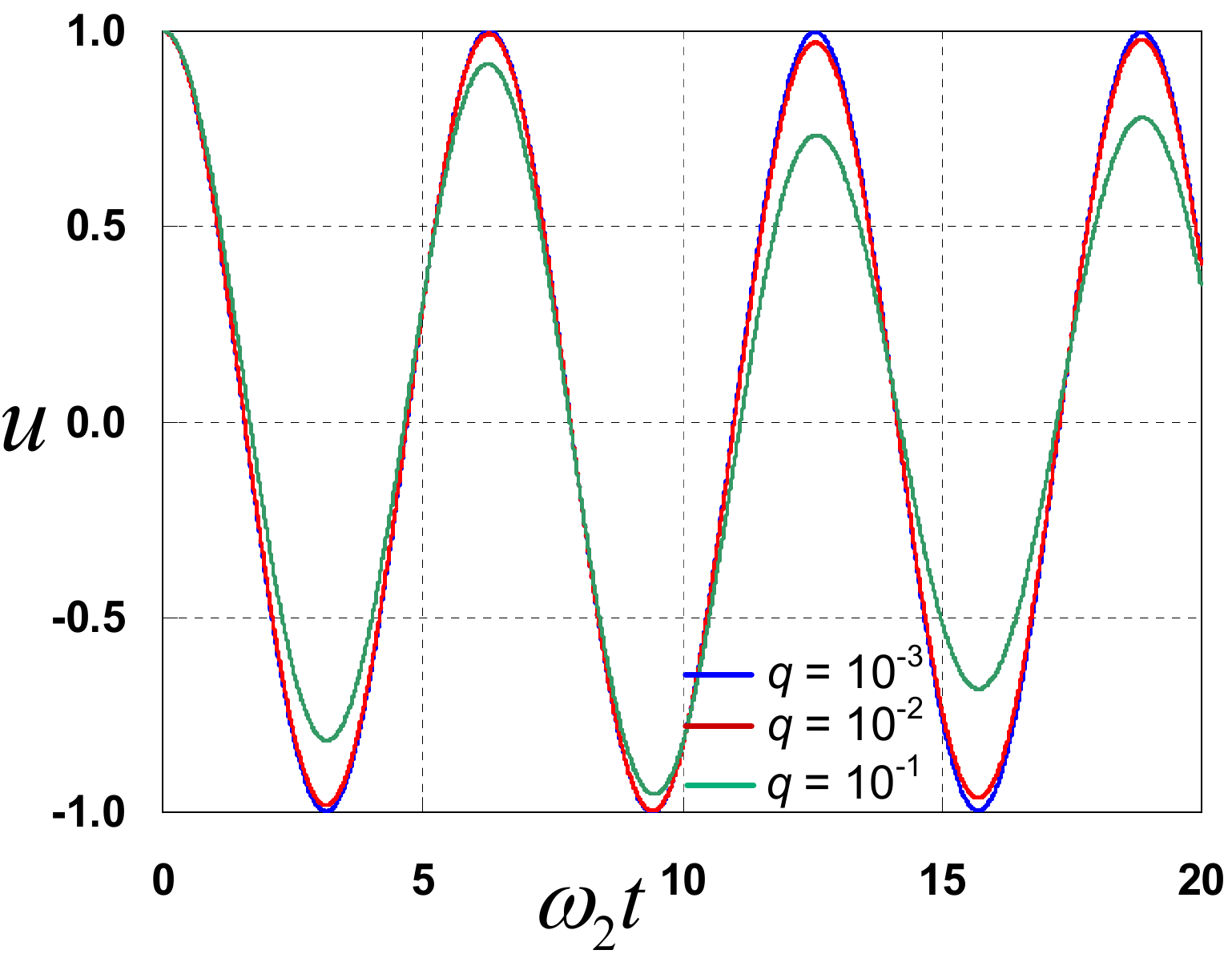}
	 \label{disp_sphere_mode_2}
	}
\subfigure[third mode]
 {
     \includegraphics[scale = 0.43]{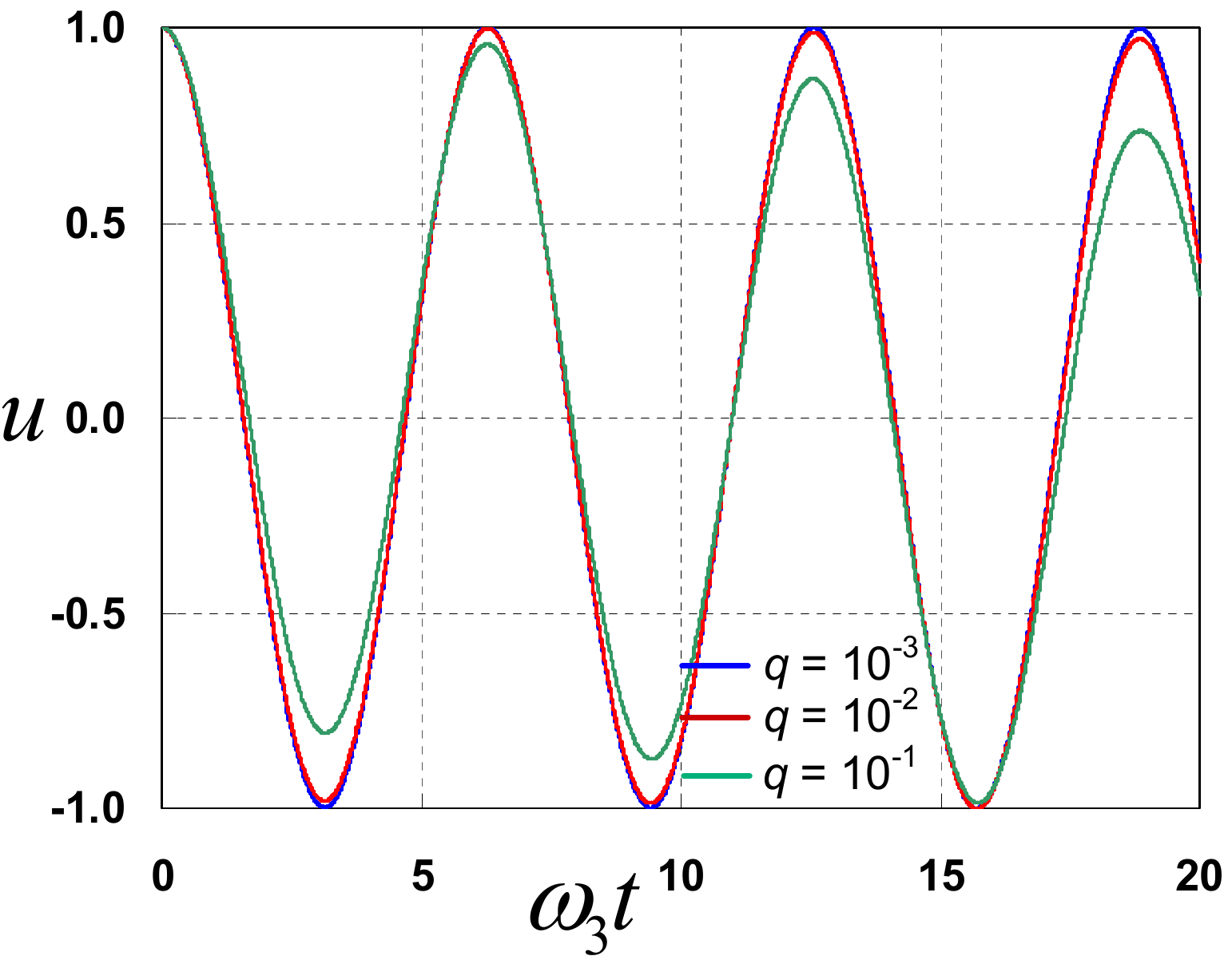}
	 \label{disp_sphere_mode_3}
	}	
\caption{Transient solutions of the sphere's outer surface normalized displacement for the first three modal excitations.}
\label{Sphere_Disp_Figs}
\end{figure}

\begin{figure}
\centering
\scalebox{0.6}
{
	\includegraphics{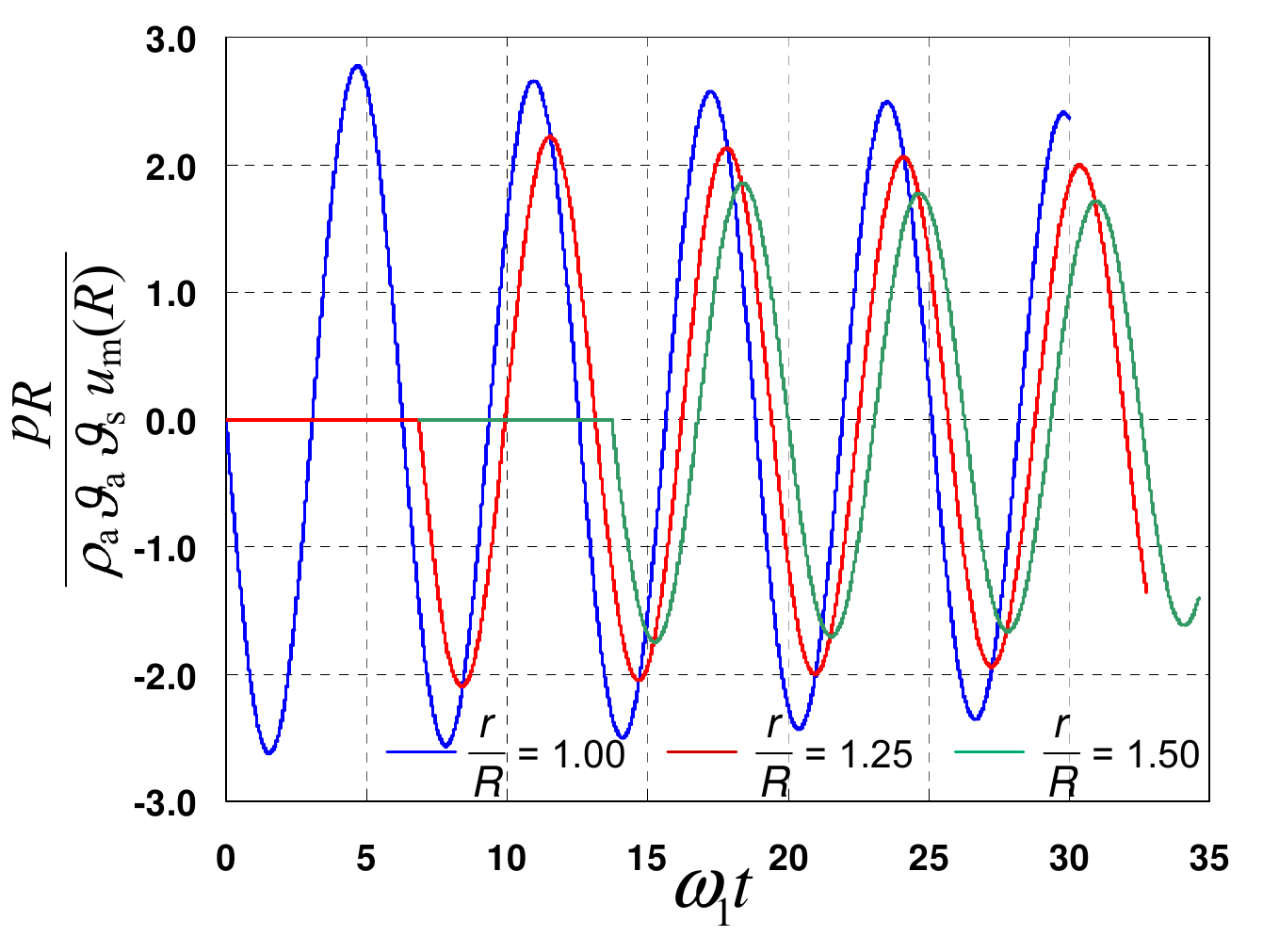}
}
\caption{Transient solution of the normalized acoustic pressure at three different radial locations for the first modal excitation, $q = 10^{-2}$, $\frac{\vartheta_\textnormal{a}}{\vartheta_\textnormal{s}} = 0.1$}
\label{Sphere_Pressure_Figs}
\end{figure}

The displacement $u(r,t)$ is obtained by inverting the Laplace expression of Eq. \eqref{U_set_eq}. As such, we have
\begin{equation}
u(r,t) = \cos(t) \ast u_1(t) \ast u_2(r,t) + \dfrac{u_\textnormal{m}(r)}{u_\textnormal{m}(R)}\cos(t)
\end{equation}
where $\ast$ refers to the convolution operator. On the outer surface, the displacement simplifies to
\begin{equation}
	\label{out_surf_disp_response}
	u(R,t) = \cos(t) \ast u_1(t) + \cos(t)
\end{equation}
The Laplace inversion is performed using the Bromwich method explained in appendix B.2 of \cite{Graff_book} and chapter 2 of \cite{Cohen_book}. In the general case, we have
\begin{equation}
\label{Res_eq}
u(t) = \dfrac{1}{2 \pi i} \int\limits_{\gamma - i\infty}^{\gamma + i\infty} \exp\left(st\right)\hat{U}\left(s \right) \textnormal{d}s = \sum\limits_{k=1}^{N_p} \textnormal{Res}\left(\hat{U}\left(s\right), s_k\right)\exp\left(s_kt\right)
\end{equation}
where $s_k$ is the $k$th pole of $\hat{U}(s)$ and $N_p$ is the total number of poles. In this problem, $\hat{U}_1(s)$ has infinitely many poles (one is real and the remaining are complex conjugates), thus if the sum in Eq.~\eqref{Res_eq} were to converge, it can be truncated. Asymptotic analysis shows that the complex part of the poles increases indefinitely rendering their residues insignificant. We can therefore safely disregard the contribution of these large poles and consider the first pole pairs with the smallest complex part. We have indeed considered the first fifty conjugate pairs of poles. The same procedure applies to $\hat{U}_2(r,s)$. The poles were obtained numerically via the two-dimensional bisection method (a root finding algorithm of assured convergence). The convolution integral is performed numerically using the cubic splines method. The displacement response at the outer surface for the first three modes at three different values of $q$ is shown in Fig.~\ref{Sphere_Disp_Figs}. For real applications, the solid is primarily metallic and the fluid is ambient air; thus a $q$ value of $10^{-4} \sim 10^{-3}$ is reasonable. For undersea applications, $10^{-1}$ would be a realistic value for $q$. But in this case, the viscous dissipation in water, especially at low speed (low Reynolds number), would intensify, spoiling the major assumption of its negligence when evaluating acoustic damping. In any case, the system is simulated at $q \in \{10^{-3},10^{-2},10^{-1}\}$ in order to test the validity of the $\zeta_{\textnormal{m,f}}$ expression over a wide range of $q$. In all cases, we fixed $\frac{\vartheta_\textnormal{a}}{\vartheta_\textnormal{s}} = 0.1$. The transient solution of the acoustic pressure at three spots in the fluid domain is shown in Fig.~\ref{Sphere_Pressure_Figs}.

\begin{figure}
\centering
\subfigure[$q = 10^{-3}$]
	{
	\includegraphics[scale = 0.42]{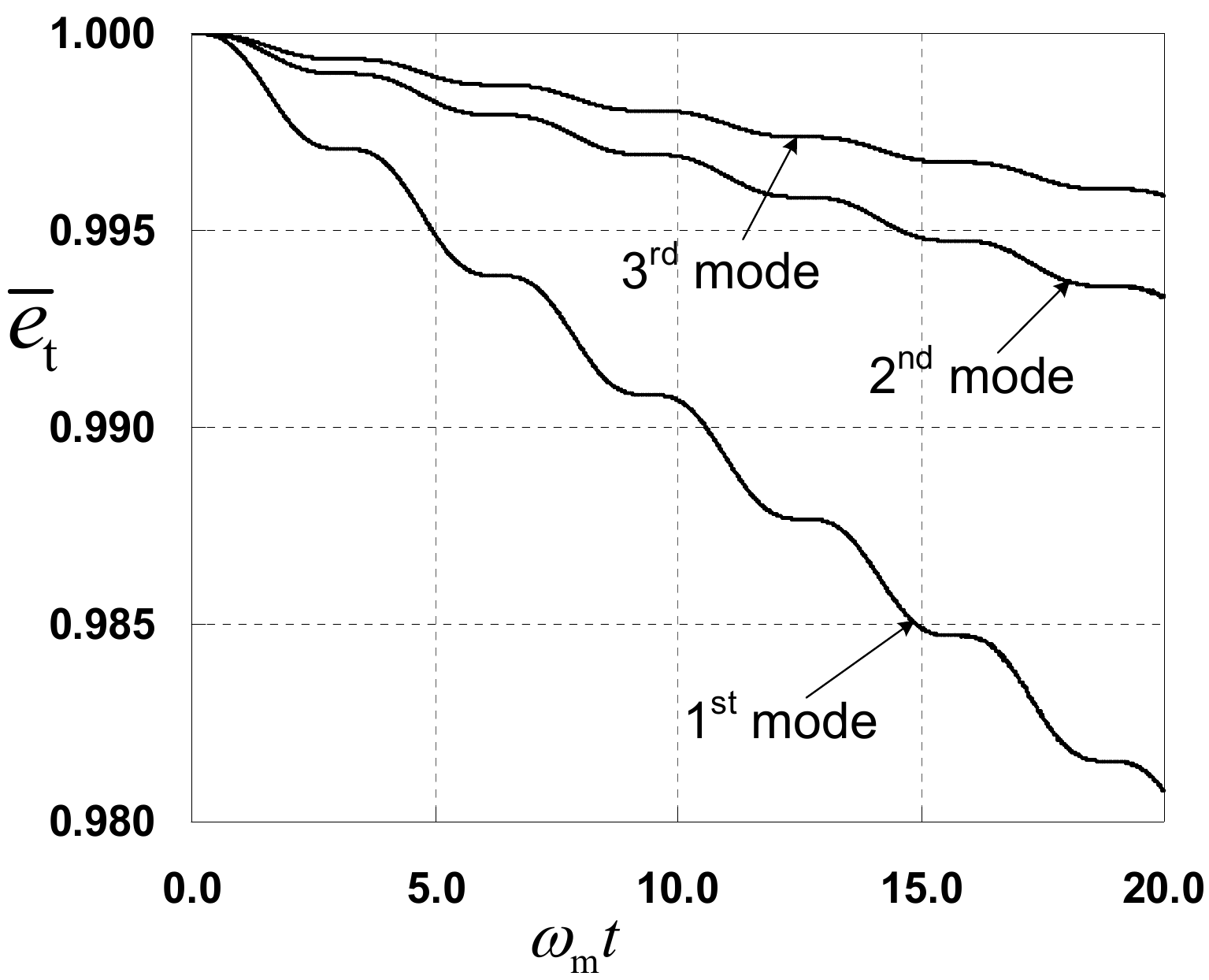}
	\label{ene_sphere_q_1}
}
\subfigure[$q = 10^{-2}$]
 {
     \includegraphics[scale = 0.42]{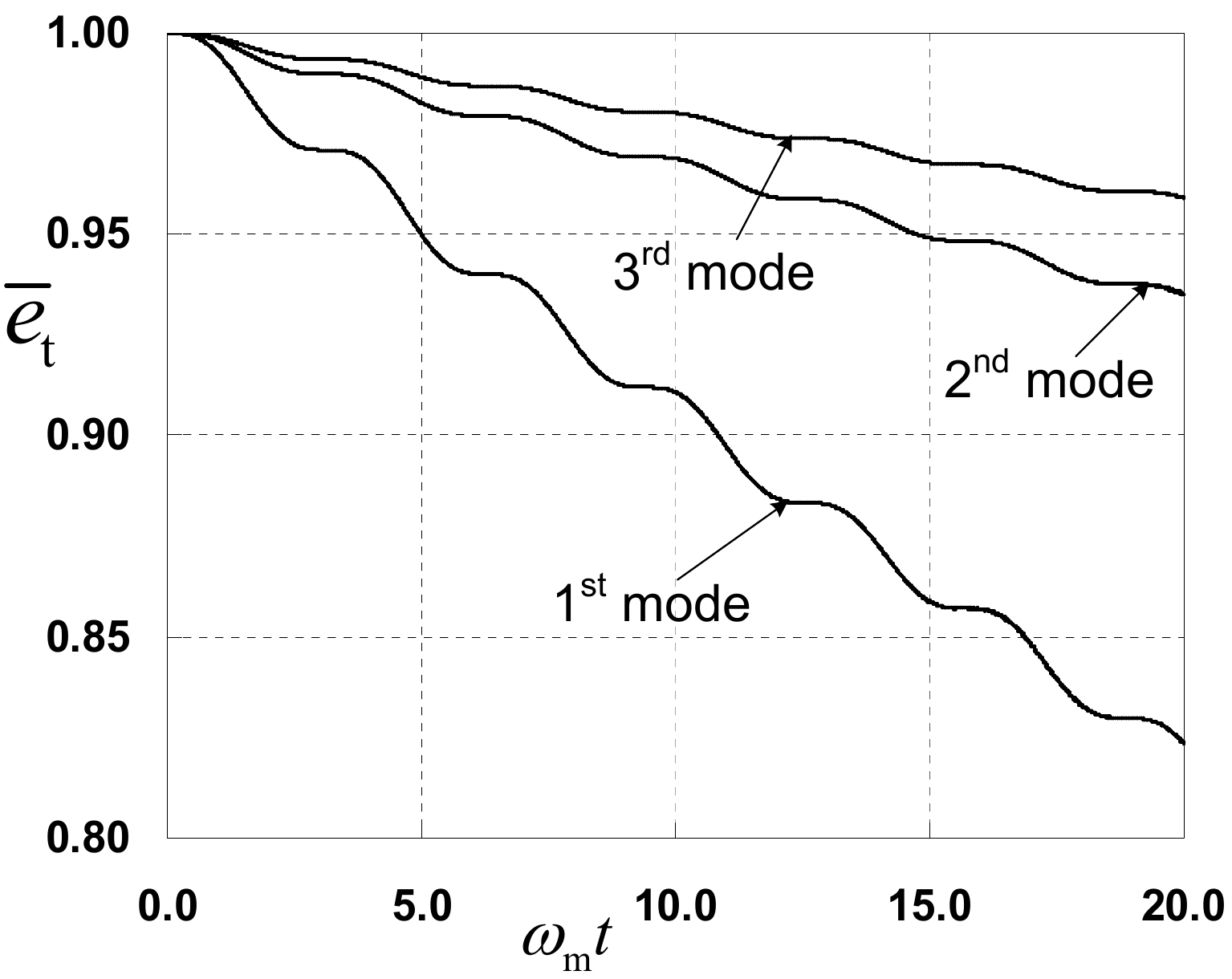}
	 \label{ene_sphere_q_2}
	}
\subfigure[$q = 10^{-1}$]
 {
     \includegraphics[scale = 0.42]{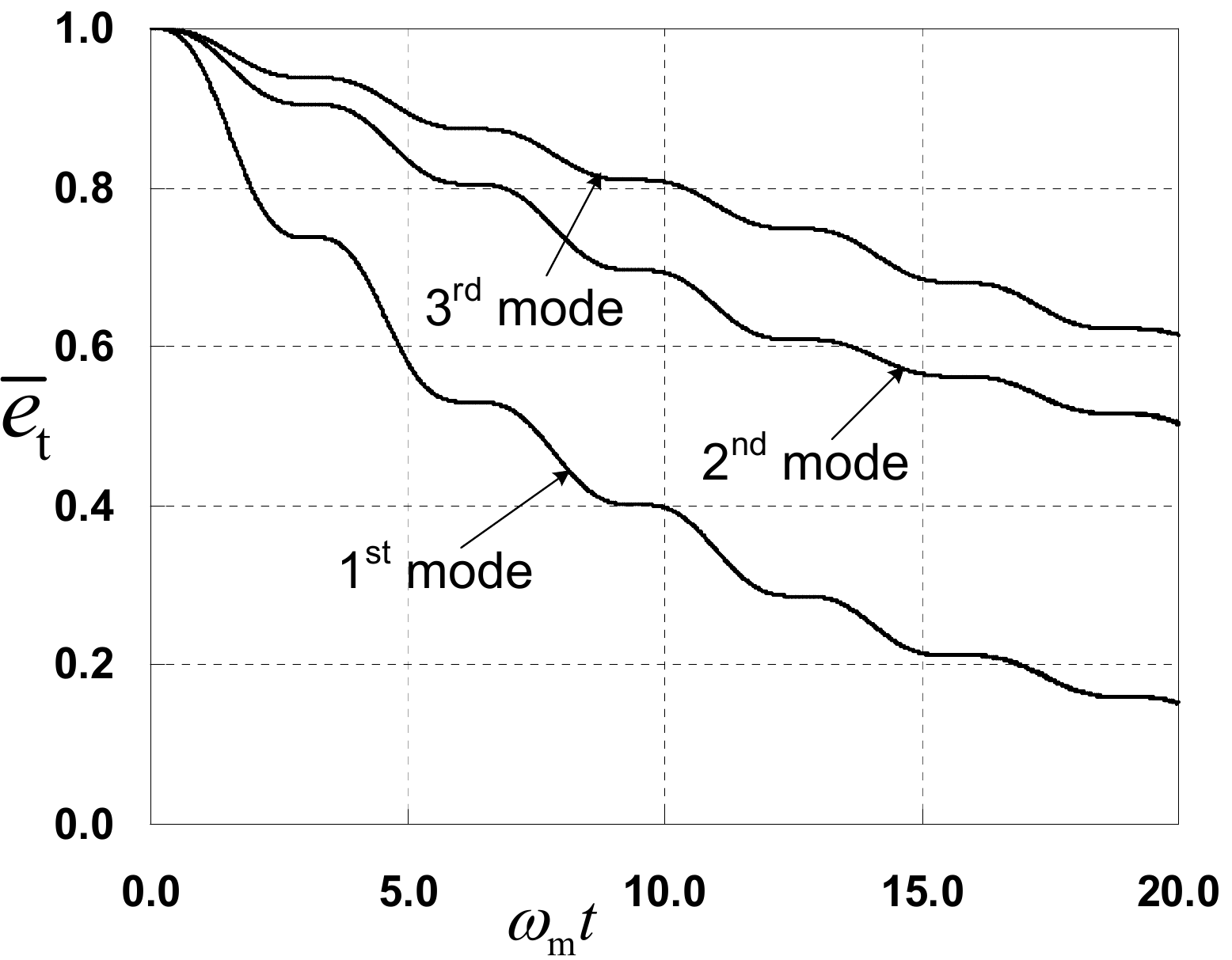}
	 \label{ene_sphere_q_3}
	}
\caption{Transient solution of the sphere's volume averaged normalized total energy}
\label{Sphere_Ene_Figs}
\end{figure}

The transient response due to the first modal excitation reveals damping characteristics similar to those of the harmonic oscillator where the magnitude of the extrema points decay steadily with time. However, for the second and higher modes, this steady decay property is lost. This observation is justified by the presence of the secondary (minor) peaks in the frequency response shown in Fig. \ref{Bode_sphere_plot}. The presence of these minor peaks points to a slight contribution into the transient response from an additional frequency, the issue that leads to an intermittent perturbation of the steady decay of the values of the extrema points. Being the case, the displacement version of the log decrement method fails to predict the damping ratio for other than the first mode; the energy version comes into effect to fulfil this prediction. The strain field, along with the velocity field are obtained from the displacement solution via direct numerical differentiation with respect to space and time. The strain and the kinetic energy are consequently evaluated in conjunction with Eq.~\eqref{average_strain_energy_Eq} and Eq.~\eqref{average_kinetic_energy_Eq}. The total energy is then obtained for the first three modes at the three chosen values of $q$ as shown in Fig.~\ref{Sphere_Ene_Figs}. The energy-based log decrement method is finally applied and the extracted damping coefficients are listed in Table ~\ref{Damping_Table}. Strong agreement between the various coefficients is observed particularly at small $q$ and low modes.

\begin{table}
\caption{Summarized damping evaluation results. Clear matching is noted at low $q$}
\begin{center}
\label{Damping_Table}
\begin{tabular}{l l c c l }
\hline
                      & $q$          & $\zeta_{\textnormal{f}}$ \%  & $\bar{\zeta}_{\textnormal{dis}}$ \% & $\bar{\zeta}_{\textnormal{en}}$ \%  \\
\hline
first mode       & $10^{-3}$ & 0.049               & 0.049                  & 0.049            \\
                      & $10^{-2}$ & 0.496               & 0.491                  & 0.494             \\
                      & $10^{-1}$ & 4.942               & 4.816                  & 4.849             \\
\hline
second mode & $10^{-3}$ & 0.017               & NA                      & 0.017              \\
                      & $10^{-2}$ & 0.173               & NA                      & 0.171              \\
                      & $10^{-1}$ & 1.726               & NA                      & 1.730              \\
\hline
third mode      & $10^{-3}$ & 0.010               & NA                      & 0.011              \\
                      & $10^{-2}$ & 0.110               & NA                      & 0.108              \\
                      & $10^{-1}$ & 1.098               & NA                      & 1.191              \\
\hline
\end{tabular}
\end{center}
\end{table}

\section{Radiation Resistance Analysis}
In this section, we conduct a qualitative comparison with regard to acoustic damping characteristics between the coupled systems discussed in this paper and some other previously analysed ones. Indeed, a significant resemblance is noticed between the findings, validating the analysis performed in this work and corroborating the dependence of acoustic damping on the physical properties of the coupled system. The key parameter on which the following analysis is centred is the acoustic radiation resistance $R_\textnormal{rad}$, which is nothing but the equivalent of the damper in the under-damped harmonic oscillator.

In \cite{Lyon_book}, an expression for the acoustic radiation resistance of a one-sided flat plate subject to excitation at frequency $f$ higher than a certain critical frequency $f_\textnormal{c}$, is given by
\begin{equation}
\label{Rad_Res_plate_eq}
R_\textnormal{rad,plate} = \frac{\rho_\textnormal{a} \vartheta_\textnormal{a} A}{\sqrt{1 - \frac{f_\textnormal{c}}{f}}} \quad \textnormal{for} \; \; f>f_\textnormal{c}
\end{equation}
For our problems, the radiation resistance is obtained through the following evaluation
\begin{equation}
\label{Rad_Res_eq}
R_\textnormal{rad} = 2 m \zeta \omega 
\end{equation}
Applying Eq.~\eqref{Rad_Res_eq} on the spherical systems, we obtain after simplifying the expressions of $\zeta$ by suppressing higher order terms
\begin{subequations}
\begin{equation}
\label{Rad_Res_thin_eq}
R_\textnormal{rad,shell} = \rho_\textnormal{a} \vartheta_\textnormal{a} A
\end{equation}
\begin{equation}
\label{Rad_Res_sph_eq}
R_\textnormal{rad,sphere} =  \frac{2}{3} \rho_\textnormal{a} \vartheta_\textnormal{a} A
\end{equation}
\end{subequations}
where $A$ is the surface area in contact with the acoustic medium. The direct dependence of the resistance value on the $\rho_\textnormal{a} \vartheta_\textnormal{a} A$ factor is noticed in the spherical problems as is the case with the flat plate one. As such, we conjecture that acoustic damping in both natural and forced excitations for general structures is dominated by this factor. For the case of forced excitations, frequency effects arise but can be lumped in a secondary proportionality factor as shown in Eq.~\eqref{Rad_Res_plate_eq}.

\section{Conclusion}
Acoustic damping is demonstrated in ideal spherical structural-acoustic models and an analytical expression for this damping is formulated and verified. The under-damped linear harmonic oscillator accurately depicts the energy dissipation in the coupled system. The matching of damping coefficients obtained by various damping extraction methods corroborates the accuracy of the analysis. The availability of analytical solutions in both solid and acoustic domains permits the efficient validation of expensive FSI simulations while considering a single problem. Finally, the future trend of this work is portrayed in two courses; first is the exploration of the modal superposition effects on acoustic damping, and second is the prediction of the coupled system forced response when subject to both acoustic and structural loading. These crucial studies, if integrated to the work achieved in this paper, provide a considerable advancement in understanding real multi-physics problems with ultimate aim to simplify the design procedures of FSI applications.

\subsection*{Acknowledgments}
This research was made possible by the support from Sandia National Laboratories -- DTRA (grant HDTRA1-08-10-BRCWMD) and the NSF (grant CMMI-1030940). The author thanks Dr. C.A. Morales for his support in publishing this manuscript.


\end{document}